\documentclass[fleqn,usenatbib]{mnras}

\usepackage[T1]{fontenc}

\DeclareRobustCommand{\VAN}[3]{#2}
\let\VANthebibliography\thebibliography
\def\thebibliography{\DeclareRobustCommand{\VAN}[3]{##3}\VANthebibliography}

\usepackage{graphicx}	
\usepackage{amsmath}	
\usepackage{newtxtext,newtxmath}

\title[SILCC-Zoom: dynamic balance in MC substructures]{SILCC-Zoom: the dynamic balance in molecular cloud substructures}

\author[S. Ganguly et al.]{
Shashwata Ganguly,$^{1}$\thanks{E-mail: ganguly@ph1.uni-koeln.de}
S. Walch,$^{1,2}$
S. D. Clarke$^{1,3}$
and D. Seifried$^{1,2}$
\\
$^{1}$I. Physikalisches Insitut, Universit{\"a}t zu K{\"o}ln, Z\"ulpicher Str. 77, 50937 K\"oln, Germany\\
$^{2}$Cologne Centre for Data and Simulation Science, University of Cologne, Cologne, Germany\\
$^{3}$Academia Sinica, Institute of Astronomy and Astrophysics, Taipei, Taiwan\\
}

\date{Accepted XXX. Received YYY; in original form ZZZ}

\pubyear{2022}

\begin{document}
\label{firstpage}
\pagerange{\pageref{firstpage}--\pageref{lastpage}}
\maketitle

\begin{abstract}
How molecular clouds fragment and create the dense structures which go on to form stars is an open question. We investigate the relative importance of different energy terms (kinetic, thermal, magnetic, and gravity - both self-gravity and tidal forces) for the formation and evolution of molecular clouds and their sub-structures based on the SILCC-Zoom simulations. These simulations follow the self-consistent formation of cold molecular clouds down to scales of 0.1 pc from the diffuse supernova-driven interstellar medium in a stratified galactic disc. We study the time evolution of seven molecular clouds (five with magnetic fields and two without) for 1.5-2 Myr. 
Using a dendrogram, we identify hierarchical 3D sub-structures inside the clouds with the aim to understand their dynamics and distinguish between the theories of gravo-turbulent fragmentation and global hierarchical collapse. 
The virial analysis shows that the dense gas is indeed dominated by the interplay of gravity and turbulence, while magnetic fields and thermal pressure are only important for fluffy, atomic structures.
Over time, gravitationally bound sub-structures emerge from a marginally bound medium (viral ratio $1 \leq \alpha_{\rm vir}^{\rm vol} <2$) as a result of large-scale supernova-driven inflows rather than global collapse. 
A detailed tidal analysis shows that the tidal tensor is highly anisotropic. Yet the tidal forces are generally not strong enough to disrupt either large-scale or dense sub-structures but cause their deformation. By comparing tidal and crossing time scales, we find that tidal forces do not seem to be the main driver of turbulence within the molecular clouds.
\end{abstract}




\section{Introduction}
Star formation occurs inside cold, dense and mostly molecular gas in the interstellar medium (ISM) - in so-called molecular clouds (MCs). They are typically thought of as the largest structure that can decouple from galactic dynamics \citep[][]{chevance_lifecycle_2020}. The nature of MCs, their lifetime, formation mechanism, and whether they are bound or not, all have direct consequences on our understanding of the possible ways in which dense structures and eventually stars can form inside these clouds. \par 
Several key discoveries over the past decades have facilitated and developed our understanding of molecular clouds. We know that they are in rough energy equipartition between kinetic and potential energy \citep[e.g. see][]{dame_largest_1986, solomon_mass_1987, blitz_giant_2007}.
Only a small fraction of gas inside these clouds eventually ends up in stars \citep[][]{ kennicutt_global_1998, genzel_study_2010, krumholz_star_formation_law_2012}, which leads to a low star formation efficiency of the order of a few percent, although with some variation \citep[][]{murray_star_2011}. Molecular clouds are known to exhibit supersonic linewidths \citep[][]{larson_turbulence_1981, orkisz_turbulence_2017}, but generally lack a clear collapse signature such as redshifted lines \citep[][]{zuckerman_models_1974}. While this has recently been challenged with observations of gravity driven flows \citep[][]{kirk_filamentary_2013, peretto_sdc13_2014, chen_filamentary_2019, shimajiri_probing_2019}, it is unclear if such flows extend to entire clouds \citep[][]{chevance_life_2022}. The linewidth is related to the mass and size of the molecular clouds themselves \citep[][]{larson_turbulence_1981, solomon_mass_1987, brunt_large-scale_2003}. The original power-law exponents proposed by \citet{larson_turbulence_1981} have since been revised and challenged. For example, \citet{heyer_re-examining_2009} show that the Larson coefficient has a dependency on the MC surface density, and there have been studies pointing out that the Larson relation may not hold inside a single cloud \citep[][]{wu_properties_2010, traficante_testing_2018}. \par 
There are several possible ways to interpret these key observations. The traditional view, originally proposed by \citet{zuckerman_models_1974}, is that the supersonic linewidths in observed clouds are caused by small scale turbulence that provides overall support and prevents the entire cloud from freely collapsing under gravity. In this so-called gravo-turbulent (GT) scenario, clumps and cores inside molecular clouds form as a result of compression in the highly turbulent gas \citep[see e.g.][]{klessen_formation_2001, padoan_star_2011}. Only a few of the compressed regions become gravitationally bound structures, which collapse and form stars. In this scenario, the Larson-like power law connecting the turbulent velocity dispersion, $\sigma_{v}$, and the linear size of the cloud, $R$, i.e. $\sigma_{v} \propto R^{0.5}$, originates from a turbulent cascade of large-scale motions \citep[see e.g. excellent reviews by][]{mac_low_control_2004, ballesteros-paredes_molecular_2007, mckee_theory_2007, hennebelle_turbulent_2012}.\par 
This view has since been challenged. Magnetic fields have been proposed as a possible mechanism to provide support against gravitational collapse \citep[][]{mouschovias_nonhomologous_1976-2, mouschovias_nonhomologous_1976-1, padoan_star_2011, federrath_star_2012, ibanez-mejia_gravity_2022}. However, after the emergence of key observations suggesting that most cloud structures are magnetically supercritical , it is generally accepted that magnetic fields cannot be solely responsible for the observed, low star formation efficiency \citep[][]{crutcher_magnetic_1999, bourke_new_2001, troland_magnetic_2008, falgarone_cn_2008, crutcher_magnetic_2010}. \par 
It is unclear if MCs need any support at all. Several studies suggest that MCs are transient entities, and that their lifetimes are short enough that they may not need any support \citep[][]{bash_galactic_1977, leisawiz_CO_1989, elmegreen_star_2000, hartmann_rapid_2001, engargiola_molecular_2003, kawamura_second_2009, meidt_GMC_2015, mac_low_fast_2017}. \par 
In recent years, chaotic, hierarchical gravitational collapse has emerged as another possible alternative to the gravo-turbulent scenario. The hierarchical collapse scenario, originally developed by \citet{hoyle_fragmentation_1953} and suggested as a mechanism for the collapse of a cloud by \citet{goldreich_molecular_1974}, has been considered briefly over the past decades \citep[e.g.][]{elmegreen_star_2000, hartmann_rapid_2001, vazquez-semadeni_molecular_2007}. In its most recent form, the so-called global hierarchical collapse (GHC) scenario \citep[see e.g.][]{vazquez-semadeni_global_2019}, the interstellar turbulence produces nonlinear density fluctuations, which facilitate a multi-scale hierarchical collapse of the entire molecular cloud through a mass cascade across scales. The chaotic nature of the collapse masks any obvious signatures of global collapse, and the low efficiency of star formation is explained by the fact that stellar feedback from the first generation of massive stars disperses the cloud before it can convert more of its gas mass into stars \citep[see e.g.][]{vazquez-semadeni_high-_2009, vazquez-semadeni_hierarchical_2017, ballesteros-paredes_gravity_2011, kuznetsova_signatures_2015, kuznetsova_kinematics_2018}.\par
Most analyses of the evolution of MCs focus on the self-gravity of the gas itself \citep[][]{dobbs_why_2011}. The complex nature of the medium, however, implies that tidal forces can have a potentially significant sway over the gas dynamics. For example, stability produced by tidal forces has been invoked to explain the peak of the initial mass function on core-scales \citep[][]{lee_stellar_2018, colman_origin_2020}. For star clusters, it has been shown that fully compressive tides can slow down the dissolution of clusters \citep[][]{renaud_evolution_2011}, and that young clusters are preferentially located in fully compressive regions of galaxies \citep[][]{renaud_fully_2009}. We aim to consider tidal forces here in the context of the (internal) dynamics of MCs. \par 
In the presented work, we perform a detailed energy analysis of self-consistently formed MCs and their sub-structures based on the SILCC-Zoom simulations \citep{seifried_silcc-zoom_2017}. The MCs are situated inside a larger scale multi-phase ISM environment \citep{walch_silcc_2015, girichidis_silcc_2016}. This allows us to do a rigorous analysis of the MC dynamics and to test whether the GT or the GHC scenario are in control. We apply a three-dimensional dendrogram algorithm to define hierarchical structures \citep{rosolowsky_structural_2008}.
 \par 
The paper is structured as follows. We first present the details of the numerical setup in section \ref{sec:methods}. In section \ref{sec:scaling}, we highlight the different scaling relations (Larson's velocity-size and  mass-size relations, and the Heyer relation). We then discuss the energetic balance between self-gravity, kinetic, magnetic, and thermal energy in section \ref{sec:energetics}. By studying the dynamic balance of these structures, we find that larger scale and denser, molecular structures are primarily governed by the interaction of the turbulent kinetic energy and self-gravity, with magnetic and thermal energies playing a subservient role. In section \ref{sec:gravity}, we discuss the relative importance of self-gravity and the gravity from the surrounding medium, by both a direct comparison of acceleration terms as well as a more detailed tidal analysis. From the direct comparison of the acceleration vectors, we find that bound and marginally bound structures are dominated by self-gravity over external gravity. We further assess the nature and extent of deformation caused by the gravity of the surrounding clumpy medium by performing a tidal analysis, and find that the overall gravitational field is highly anisotropic but still mostly fully compressive. Finally, we present the summary of our findings in section \ref{sec:conclusions}. 

\section{Numerical methods and simulation} \label{sec:methods}

We present results based on the SILCC-Zoom simulations \citep[][]{seifried_silcc-zoom_2017}. The SILCC-Zoom simulations are zoom-in simulations of MCs forming self-consistently within the larger-scale, multi-phase ISM SILCC simulations \citep{walch_silcc_2015, girichidis_silcc_2016}. In this section, we highlight some key features of the simulations. Further details regarding the simulations can be found in \citet{seifried_silcc-zoom_2017}. For description of the hydrodynamic (HD) and magnetohydrodynamic (MHD) clouds used for the present analysis, see \citet{seifried_silcc-zoom_2017} and \citet{seifried_silcc-zoom_2019}, respectively.\par
All simulations were performed using the adaptive mesh refinement (AMR) code FLASH, version 4 \citep[][]{fryxell_flash_2000, dubey_challenges_2008-1}. We present results from runs with and without magnetic fields. The hydrodynamic cloud simulations have been performed using the MHD 'Bouchut 5-wave solver' \citep[][]{bouchut_multiwave_2007, waagan_positive_2009}, with magnetic field strength set to zero. The MHD simulations have been run using an entropy-stable solver which guarantees positive pressure and minimizes dissipation \citep[][]{derigs_novel_2016, derigs_novel_2017,derigs_ideal_2018}.\par
The entire simulation domain consists of a box of size 500 pc $\times$ 500 pc $\times$ $\pm$ 5 kpc, with the long axis representing the vertical $z-$direction of the stratified galactic disc. The box is set with periodic boundary conditions in the $x-$ and $y-$directions, and outflow boundary conditions in $z-$direction. The initial gas surface density is set to $\Sigma_{\rm gas} = 10 \, \rm{M_{\odot}\,pc}^{-2}$ which corresponds to solar neighbourhood conditions. The vertical distribution of the gas is modelled as a Gaussian, i.e. $\rho = \rho_0 \exp(-z^2/2h_{z}^2)$, where $h_z$ = 30 pc is the scale height and $\rho_0 = 9 \times 10^{-24}$ g cm$^{-3}$ is the midplane gas density. The initial gas temperature is set to 4500 K. For runs with magnetic fields, the magnetic field is initialized along the $x-$direction, i.e. $\mathbf{B} = (B_x,0,0)$ with $B_x = B_{x,0}\sqrt{\rho (z)/\rho_0}$ and the magnetic field strength in the midplane $B_{x,0} = 3 \mu G$. The field strength is chosen to be in accordance with recent observations \citep[e.g.][]{beck_magnetic_2013}.   \par
All simulations include self-gravity as well as an external galactic potential due to old stars. The external potential is calculated using the assumption of a stellar surface density of 
$ \Sigma_{\rm star} = 30 \, \rm{M_{\odot} \, pc}^{-2}$ with a scale height of 100 pc. The self-gravity of the gas is computed using the Octtree-based algorithm by \citet{wunsch_tree-based_2018}.  \par
Apart from the dynamics of the gas, we also model its chemical evolution using a simple non-equilibrium chemical network based on hydrogen and carbon chemistry \citep{nelson_dynamics_1997,glover_simulating_2007, glover_modelling_2010}. For this purpose, we follow the abundance of ionized, atomic, and molecular hydrogen (H$^+$, H, H$_2$), carbon-monoxide (CO), ionized carbon (C$^+$), free electrons (e$^-$), and atomic oxygen (O). At the beginning of the simulation, hydrogen is fully atomic within the disc midplane and carbon is fully ionized. The simulations also include an interstellar radiation field (ISRF), with strength of $G_0=1.7$ in Habing units \citep[][]{habing_interstellar_1968, draine_photoelectric_1978}. The ISRF is attenuated depending on the column density of the gas, thereby accounting for dust shielding as well as (self-) shielding of H$_2$ and CO. This is accounted for using the \textsc{TreeRay} \textsc{Optical Depth} module developed by \citet{wunsch_tree-based_2018} and takes into account the 3D geometry of the embedded region. \par
The turbulence in the simulations is generated by supernova explosions (SNe). The explosion rate is set to 15 SNe per Myr, which is consistent with the star formation rate surface density expected for the given gas surface density from the Kennicutt-Schmidt relation \citep[][]{schmidt_rate_1959, kennicutt_global_1998}. 50\% of the SNe are placed following a Gaussian random distribution with a scale height of 50~pc along the $z-$direction , while the other 50\% are placed at density peaks of the gas. This prescription of mixed supernova driving creates a multi-phase turbulent ISM which can be used as initial condition for the zoom-in simulations \citep{walch_silcc_2015, girichidis_silcc_2016}. \par
For the SILCC-Zoom simulations, we drive the medium with SNe up to a certain time $t_0$. Up to $t_0$, the uniform grid resolution is 3.9 pc. At $t_0$, the SNe are stopped and multiple regions are identified for the zoom-in process, primarily due to their molecular gas content at a later simulation time, when the simulation has been continued on the coarse resolution. This corresponds to the start of the evolution of the zoom-in regions, and we refer to this point as $t_{\mathrm{evol}}=0$. The total simulation time $t$ is related to the cloud evolution time $t_{\mathrm{evol}}$ as
\begin{equation}
    t = t_0 + t_{\mathrm{evol}}.
\end{equation}
The different simulations are compared at similar $t_{\mathrm{evol}}$. For tracing the evolution of the clouds, in the selected regions, the AMR grid is allowed to refine to a higher resolution in order to capture the internal structure of the forming MCs. These regions are referred to as "zoom-in regions". In each SILCC box we run two zoom-in regions simultaneously, such that we follow eight zoom-in regions in total. The different MHD runs are seeded with a different set of of random supernovae. All other initial conditions are the same. All the runs presented here have an effective spatial resolution of 0.125 pc. For more details on the zoom-in process see \citet{seifried_silcc-zoom_2017}.  \par

\subsection{Cloud selection}
\begin{figure*}
    \centering
    \includegraphics[width=\textwidth]{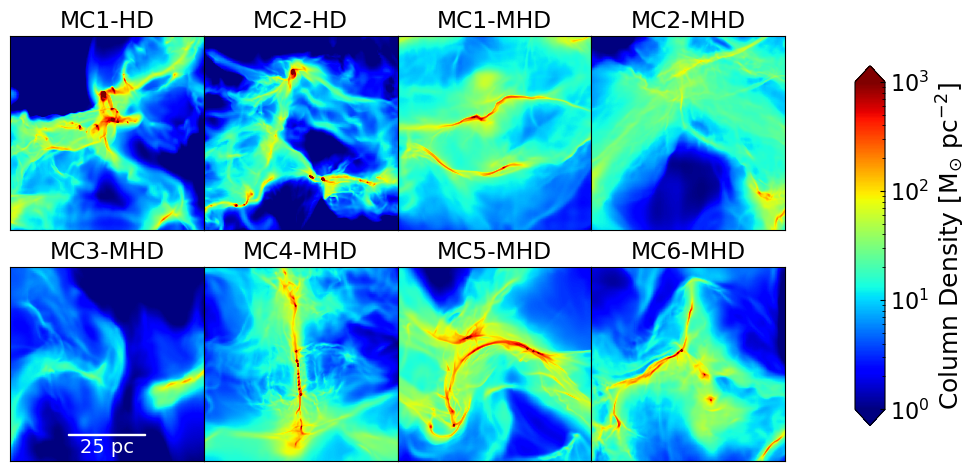}
    \caption{Column densities of the analyzed clouds, each projected along the $x-$axis at $t_{\rm evol}=$3.5 Myr. The side length of each plot is 62.5~pc.
    The MHD clouds (top right, and bottom row) have elongated filamentary structures with extended diffuse envelopes, while the hydrodynamic clouds (top left) appear to be more clumpy and fragmented. Note that MC3-MHD has been excluded in further analysis due to its lack of discerning structures and low molecular gas content (see Table \ref{tab:cloud_info}).}
    \label{fig:projection}
\end{figure*}

For the analysis presented in this work, we look at eight different (62.5 pc)$^3$ zoom-in regions. 
Each zoom-in region contains one molecular cloud (called 'MC1', 'MC2', and so on). Six simulations include magnetic fields (run names with '-MHD'), while the other two are of hydrodynamic nature ($B=0$; run names with '-HD'). We present some basic information of the different MCs in Table \ref{tab:cloud_info}, and a projected view of all eight clouds at time $t_{\mathrm{evol}}= 3.5$~Myr in Fig.~\ref{fig:projection}. The clouds we analyze have total gas masses of $5.4-10.1 \times 10^{4}$~M$_{\odot}$, and H$_2$ masses between $0.9-2.1 \times 10^{4}$~M$_{\odot}$ at t$_{\rm evol}=2$~Myr. Note that the time at which we begin the zoom-in, $t_0$, varies for the different simulations (see Table \ref{tab:cloud_info}). 
\begin{table}
\centering
    \begin{tabular}{l|c|c|c|c}
        \hline
         Run name & MHD & $t_0$  & Total mass & H$_2$ mass\\
         &   & [Myr] &$[10^4\,{\rm M}_{\odot}]$& $[10^4\,{\rm M}_{\odot}]$\\
         \hline
         MC1-HD&  no& 12& 7.3 & 2.1 \\
         MC2-HD&  no& 12& 5.4 & 1.6 \\
         \hline
         MC1-MHD& yes& 16& 7.8 & 1.3 \\
         MC2-MHD& yes& 16&  6.2 & 0.86 \\
         (MC3-MHD$^{\rm a}$& yes& 16& 2.0 & 0.19) \\
         MC4-MHD& yes& 11.5&  6.8 & 1.2 \\
         MC5-MHD& yes& 11.5&  10.1 & 1.6 \\
         MC6-MHD& yes& 16& 6.6 & 1.4  \\
         \hline
    \end{tabular}
    \caption{Basic information on the eight analyzed simulations. From left to right we list the run name, whether it is a magnetized run or not, the time when the AMR zoom-in started, as well as the total mass and the molecular hydrogen mass at $t_{\rm evol}=2$ Myr.\newline 
    $^{\rm a}$We discard MC3-MHD from our further analysis because of its low molecular gas content and lack of dense structures (see also Fig.~\ref{fig:projection}).}
    \label{tab:cloud_info}
\end{table}
The evolution time $t_{\mathrm{evol}}$, which we refer to throughout the paper when comparing the results for the different MCs, represents the "age" of the molecular cloud and corresponds to the time that has passed since the start of the zoom-in. \par
We perform a detailed analysis of the various clouds at different times between $t_{\mathrm{evol}}=1.5 - 3.5$~Myr for the two hydrodynamic clouds, as well as for $t_{\mathrm{evol}}=2 - 3.5$~Myr for the MHD clouds. We include the earlier time step for the hydrodynamic clouds as they evolve more rapidly compared to their MHD counterparts and show discernible structures early on. We do not include sink particles in these simulations, and hence the end time is chosen such that the cloud sub-structures are still well resolved with respect to the local Jeans length \citep{truelove_jeans_1997}.
Of the eight simulations, we find that even at $t_{\mathrm{evol}}=3.5$~Myr, MC3-MHD has a very low molecular gas fraction ($\sim$1900 M$_{\odot}$ molecular H$_2$ of a total mass of $\sim$20,000~M$_{\odot}$, see Table \ref{tab:cloud_info}). The resulting dendrogram analysis is rather incomplete because this cloud lacks discernible dense structures. Therefore, we omit MC3-MHD from our further analysis. 

\subsection{Sub-structure identification and modelling}\label{sec:dendrogram}
\begin{figure*}
    \centering
    \includegraphics[width=0.35\textwidth]{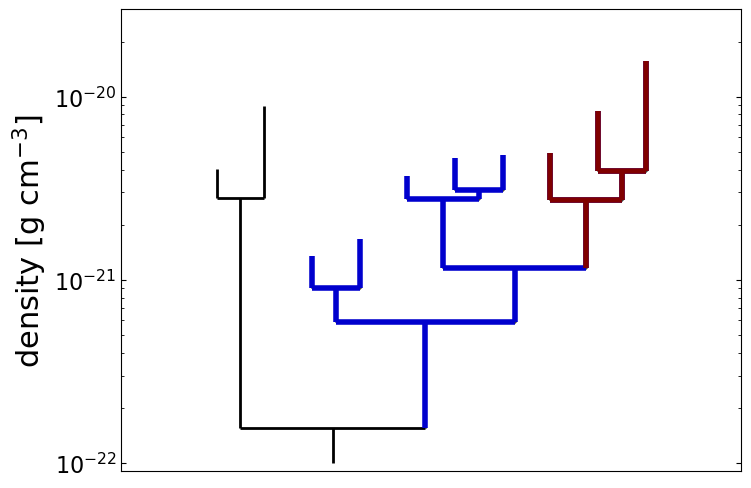}
    \includegraphics[trim=8 2 9 2, clip, width=0.64\textwidth]{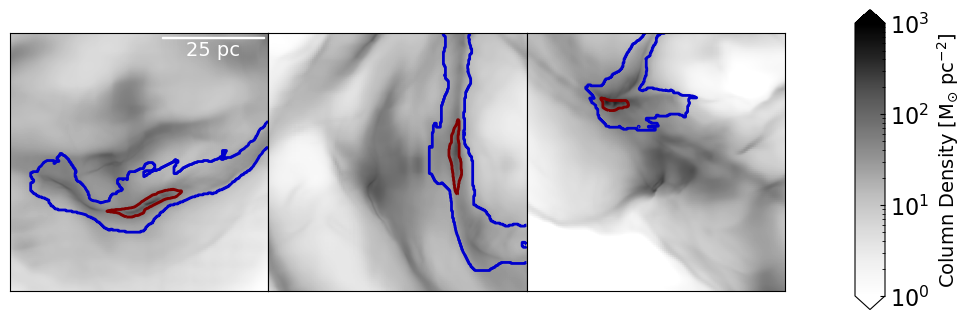}
    \caption{Left: Dendrogram tree with two sub-structures highlighted in blue and red for MC1-MHD at $t_{\rm evol}=2$~Myr, plotted using \texttt{astrodendro}. Right: We depict these two sub-structures as contours on the column density maps of MC1-MHD projected along the $x$-, $y$-, and $z$-axis, respectively. The blue and red contours correspond to the branches of the same colour from the left panel. It is apparent that the structures look filamentary on the $x$-, $y$-projections, but rather clumpy on the $z$-projection.}
    \label{fig:glue}
\end{figure*}

To identify structures in the MCs, we use a dendrogram algorithm \citep[][]{rosolowsky_structural_2008}. Dendrograms are a model independent method to determine hierarchical structures. We specifically use the python package \texttt{astrodendro} \citep[][]{robitaille_astrodendro_2019} and apply the method to find density (sub-)structures in 3D. For this purpose, we convert the AMR data to a uniform grid at our effective spatial resolution of 0.125 pc. 
Given the initial density cubes, our dendrogram analysis essentially depends on three free parameters: the initial starting threshold $\rho_0$, the density jump $\Delta \rho$, and the minimum number of cells included in a structure, $N_{\rm cells}$. For the analysis presented here, unless otherwise stated, we use $\rho_0=10^{-22}$ g cm$^{-3}$, $\Delta(\log_{10}\rho) = 0.1$ and $N_{\rm cells} = 100$. Defining $\Delta(\log_{10}\rho)$ implies that the dendrogram is built on the logarithm of the density. 
In order to reduce the number of random fluctuations, we only retain those structures in the dendrogram tree, which eventually reach a density of $\rho_{\rm peak} > 10^{-21}\,{\rm g\, cm}^{-3}$. A view of the resulting dendrogram tree can be seen in the left panel of Fig. \ref{fig:glue} for MC1-MHD at $t_{\rm evol}=2$~Myr. We highlight two branches of the dendrogram tree to show the hierarchical nature of the structures, a larger scale structure in blue, and a contained denser filamentary structure in red. The projections of these two structures can be seen in the right panel of Fig. \ref{fig:glue}, as contours over column density maps projected along the $x-$, $y-$, and $z-$direction, respectively.\par

We note that we varied the three free parameters within a reasonable range in order to test the robustness of our results. Increasing $N_{\rm cells}$ washes out the smallest sub-structures, while decreasing $\rho_0$ results in larger structures on the largest scales, but leaves the smaller scales unaffected. Increasing the minimum value of $\Delta \rho$ causes fewer structures to be identified but does not affect the general trends for different sub-structures. We have also performed the dendrogram analysis by using linear density jumps (i.e. dendrogram built on the density field instead of the logarithmic density field) of $\Delta \rho = 10^{-22}$ g cm$^{-3}$ and $\Delta \rho = 10^{-23}$ g cm$^{-3}$, in addition to using logarithmic density bins. This also leaves the statistical properties of the resulting structures basically unaffected. \par
\begin{figure*}
    \centering
    \includegraphics[width=0.99\textwidth]{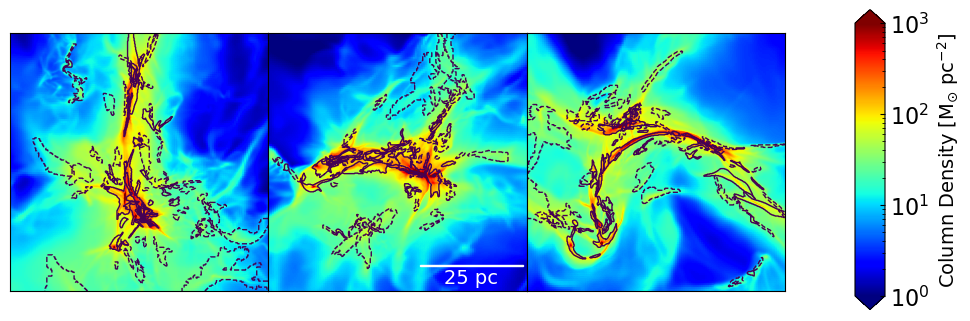}
    \caption{Left to right: Projections of MC5-MHD at $t_{\rm evol}=$3.5 Myr along the $x$-, $y$-, and $z$-axis, respectively. The contours show the projections along the same axis of the leaf dendrogram structures. Molecular structures ($> 50\% \; \mathrm{H}_2$) are plotted with solid lines, and atomic structures ($< 50\% \; \mathrm{H}_2$) are plotted with dashed lines. The molecular structures nicely trace the dense spine of the cloud, while the atomic structures mostly represent the envelope.}
    \label{fig:contours}
\end{figure*}

Examples of the resulting dendrogram structures can be seen in Fig. \ref{fig:contours}, for MC5-MHD at $t_{\mathrm{evol}}=3.5$ Myr. Projections of different dendrogram leaf structures (i.e. strutures that contain no further sub-structures inside them) are plotted over the column density as contours. We differentiate between sub-structures based on their molecular content. Molecular structures (H$_2$ mass fraction > 0.5) are plotted in solid lines, while atomic (H$_2$ mass fraction < 0.5) structures are plotted in dashed lines. Visually, we see that the molecular structures trace the denser spine of the cloud, while the atomic structures represent more the surrounding envelope. The number of total, as well as molecular structures for the various analyzed MCs can be seen in Table \ref{tab:dendro_molecular_number} for $t_{\rm evol}=3.5$~Myr. From Table \ref{tab:dendro_molecular_number}, we can see that apart from the more diffuse MC2-MHD, almost all the clouds are primarily dominated by molecular structures at the latter stage. The exact number of structures depends on the dendrogram parameters chosen, but the fraction of molecular and atomic structures is nonetheless indicative of the chemical state of the clouds. \par
\begin{table}
\centering
    \begin{tabular}{l|c|c|c|c}
        \hline
         Run name & Total & Molecular  & Molecular structure\\
         &structures &structures&fraction\\
         \hline
         MC1-HD& 214& 187 & 0.87 \\
         MC2-HD& 138& 113 & 0.81 \\
         \hline
         MC1-MHD& 52& 37 & 0.71 \\
         MC2-MHD& 127& 36 & 0.28 \\
         MC4-MHD& 172& 122 & 0.71 \\
         MC5-MHD& 208& 136 & 0.65 \\
         MC6-MHD& 120& 97 & 0.81  \\
         \hline
    \end{tabular}
    \caption{Information on the number of molecular structures obtained in the dendrogram analysis. From left to right are listed the run name, the number of total analyzed structures in each region, the number of structures which have >50\% of their hydrogen mass in molecular form, and the fraction of such molecular structures at $t_{\rm evol}=3.5$~Myr. The exact number of structures depends on the dendrogram parameters. The high fraction of molecular structures is partially a result of specifying that all structures should have a peak density $\rho_{\rm peak}>10^{-21}$~g~cm$^{-3}$.}
    \label{tab:dendro_molecular_number}
\end{table}
Once we obtain the dendrogram, we aim to approximate the density sub-structures as ellipsoids. In a companion paper (Ganguly et al., in prep.), we use this method to classify the morphology of the given structures. For the present paper, however, it allows us to compute a number of physical quantities dependent on the shortest axis of a given structure (such as the crossing time, see Fig.~\ref{fig:tratio_gext}).\par
For each structure, we compute an equivalent ellipsoid that has the same mass and the same moments of inertia as the original structure. Let us consider a uniform density ellipsoid of mass $M$ and semi-axis lengths $a, b, c$ where $a\geq b\geq c$. The moments of inertia along the three principal axes will be given as follows:
\begin{equation}
    \begin{split}
        I_a &= \frac{1}{5} M(b^2+c^2)\,\\
        I_b &= \frac{1}{5} M(c^2+a^2)\,\\
        I_c &= \frac{1}{5} M(a^2+b^2)\,
    \end{split}
\end{equation}
where $I_c\geq I_b\geq I_a$. If we now compute the principal moments of inertia of a given dendrogram structure to be $A$, $B$, and $C$, respectively, then the ellipsoid has an equivalent moment of inertia if
\begin{equation}
        A =I_a,\ B=I_b,\ C=I_c.
\end{equation}
This leads to the following equation for computing $a,\ b$, and $c$ of the equivalent ellipsoids:
\begin{equation}\label{eq:fit_ellipsoid}
    \begin{split}
        a &= \sqrt{\frac{5}{2M}(B+C-A)},\\
        b &= \sqrt{\frac{5}{2M}(C+A-B)},\\
        c &= \sqrt{\frac{5}{2M}(A+B-C)}.
    \end{split}
\end{equation}
We emphasize that the (sub-)structures identified with our dendrogram analysis represent volumina that are enclosed by the defined density contours. This does not imply that the structures are long-lived entities. They may either condense or diffuse and disperse completely. They may also accrete gas, grow in volume, merge with each other, or fragment. In the following, we analyse the dynamical state of the density (sub-)structures in detail and relate them with observed relations.
\section{scaling relations}\label{sec:scaling}

In his seminal work, \citet{larson_turbulence_1981} showed that MCs follow a relation between their linewidth, representing the 1D gas velocity dispersion, and their linear size,
\begin{equation}\label{eq:larson}
\sigma_{\rm 1D} \propto R^{n}.
\end{equation}
While \citet{larson_turbulence_1981} originally found $n=0.38$, this has later been revised to $n=0.5$ (\citet{solomon_mass_1987}, or more recently, \citet{brunt_large-scale_2003}) - roughly consistent with virialization \citep[][]{solomon_mass_1987}, turbulent cascade \citep[][]{padoan_supernova_2016}, as well as gravitational collapse \citep[][]{ballesteros-paredes_gravity_2011}. More recently, for a single molecular cloud, a much flatter velocity dispersion-size relation has been observed when studying structures with high column density. This points towards a change in the cloud dynamics once self-gravity becomes dominant \citep[][]{wu_properties_2010, traficante_testing_2018}. Re-examining Larson's laws, \citet{heyer_re-examining_2009} show that this can be explained when taking into account the surface density of the structures. Here, we investigate the Larson and Heyer scaling relations of our dendrogram structures. \par

We compute the mass-weighted 1D velocity dispersion, $\sigma_{1D}$, from
\begin{equation}\label{eq:sigma_1D}
\sigma_{\rm 1D}^2 = \frac{1}{3M}\int_V \rho (\mathbf{v} - \mathbf{v}_0)^2 \mathrm{d}^3r,
\end{equation}
 where $M$ is the total mass of the structure and $\mathbf{v}_0$  is the velocity of its centre of mass, computed as
 \begin{equation}\label{eq:v0}
 \mathbf{v}_0 = \frac{1}{M}\int_V \rho \mathbf{v} \mathrm{d}^3r.    
 \end{equation}
 The integration is performed over the entire volume $V$ of the structure. We choose a mass-weighted average because it roughly represents an intensity-weighted average, which is often used in observations. 
 We estimate the linear size $R$ from the volume of the structure as
 \begin{equation}\label{eq:size}
     R=V^{1/3}.
 \end{equation} 
 
\begin{figure*}
    \centering
    \includegraphics[trim=0 0 135 0, clip, height=4.55cm]{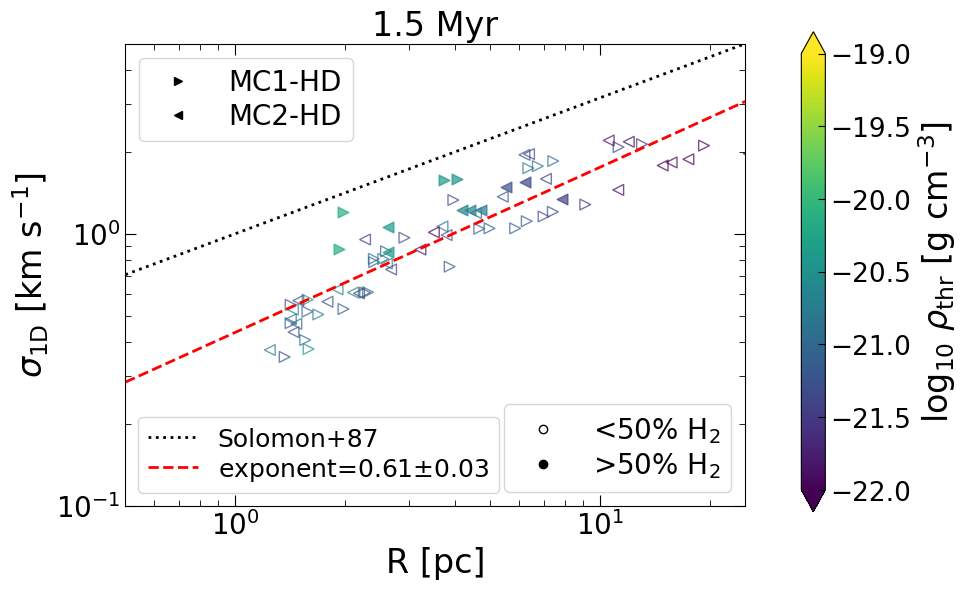}
    \includegraphics[trim=90 0 135 0, clip, height=4.55cm]{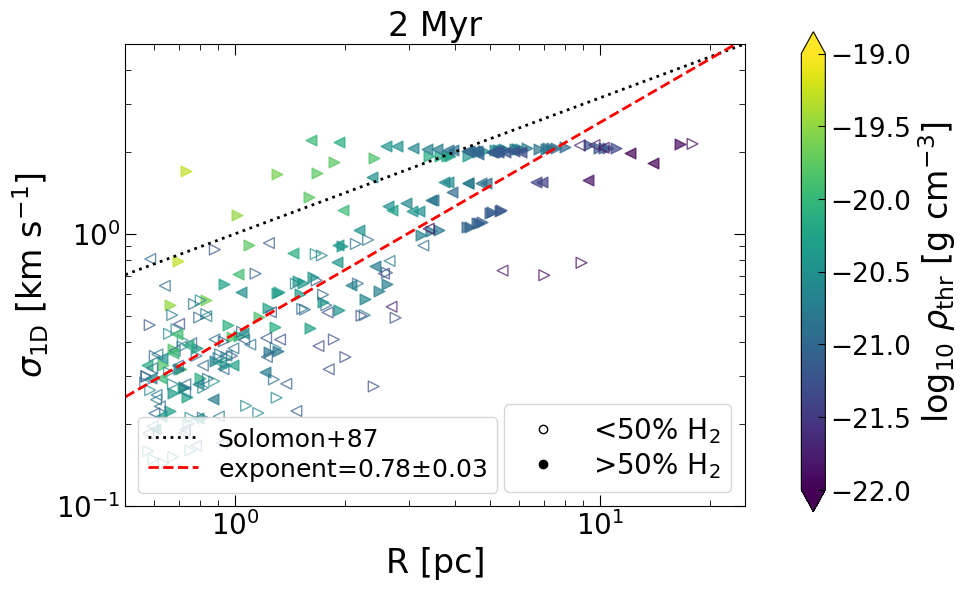}
    \includegraphics[trim=90 0 0 0, clip, height=4.55cm]{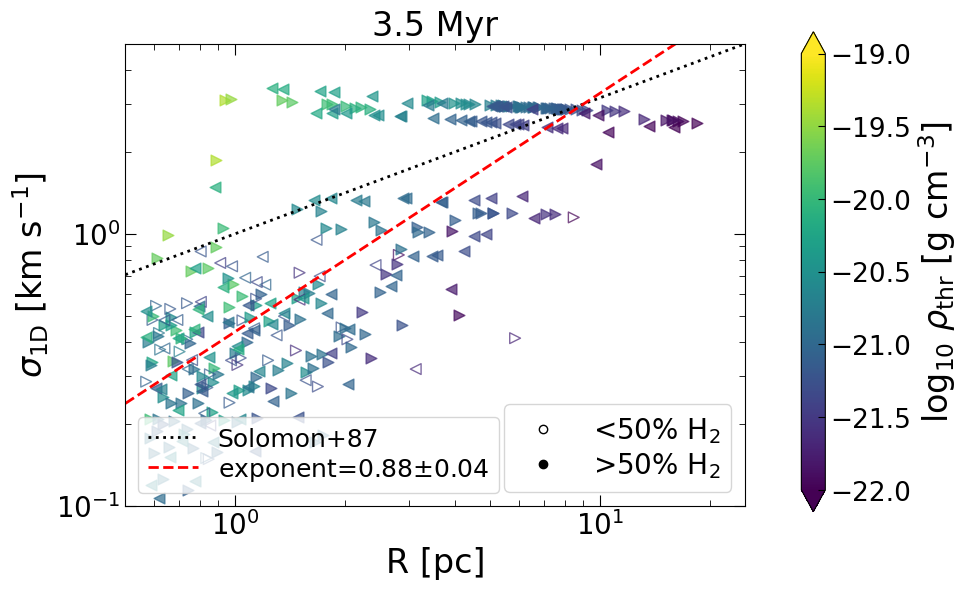}
    \includegraphics[trim=0 0 135 0, clip, height=4.55cm]{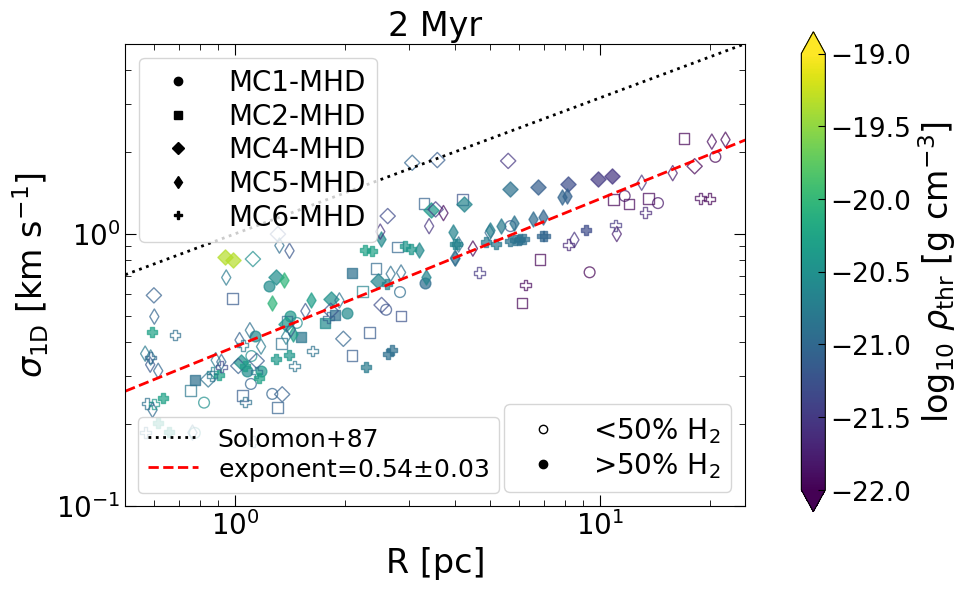}
    \includegraphics[trim=90 0 135 0, clip, height=4.55cm]{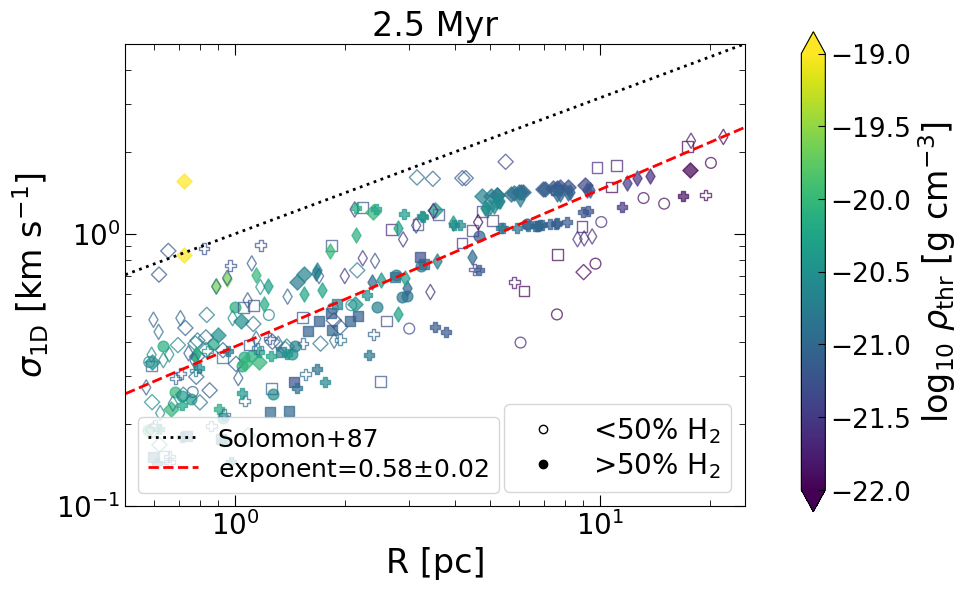}
    \includegraphics[trim=90 0 0 0, clip, height=4.55cm]{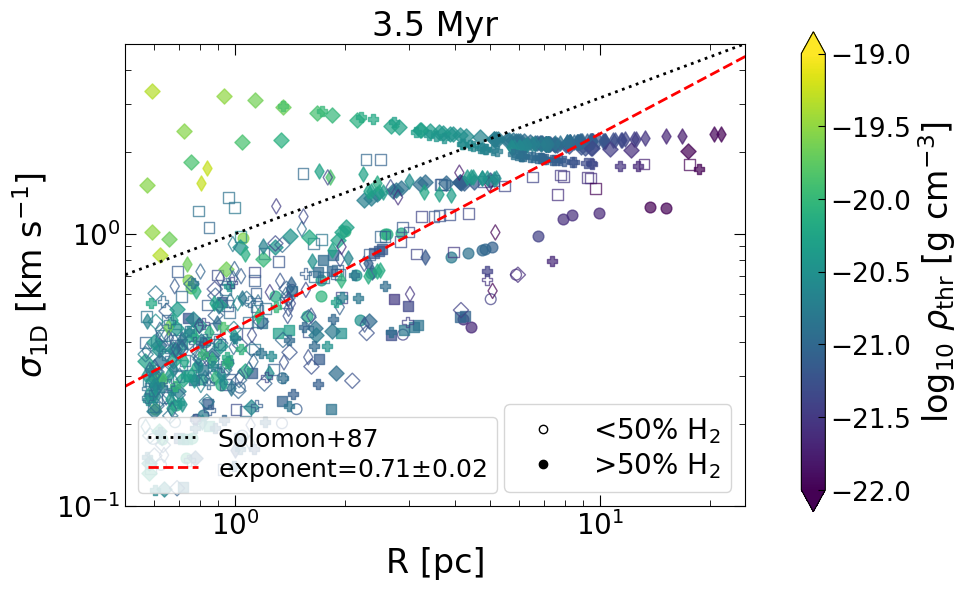}
    \caption{Time evolution of the velocity dispersion-size relation for HD clouds at 1.5, 2 and 3.5 Myr (top row, from left) and for MHD clouds at 2, 2.5 and 3.5 Myr (bottom row, from left). The colour bar represents the threshold (minimum) density inside each structure. Plotted as red, dashed and black, dotted lines are the best fit and the $\sigma \propto R^{1/2}$ obtained by \citet{solomon_mass_1987}, respectively. We show the hydrodynamic clouds at an earlier time to highlight that they behave similar to the MHD clouds. Some denser branches trail off to the left, showing complete departure from a Larson-like power law. These dense branches are absent or less pronounced at earlier times.}
    \label{fig:sigma_size}
\end{figure*}

 In Fig.~\ref{fig:sigma_size}, we plot $\sigma_{\rm 1D}$ vs. $R$ for different times (from left to right). The top row shows a compilation of all structures extracted from MC1-HD and MC2-HD, while the bottom row shows all structures found in the five MHD clouds. Note that we show slightly different times for the hydrodynamic ($t_{\mathrm{evol}}=1.5, 2.0, 3.5$~Myr) and the MHD simulations ($t_{\mathrm{evol}}=2.0, 2.5, 3.5$~Myr), owing to the fact that the former show discernible structures comparatively quicker (see below, also see \citet{seifried_silcc-zoom_2020}). We distinguish between mostly atomic structures (empty symbols) and mostly molecular structures (filled symbols) as defined in section~\ref{sec:dendrogram} (see also Fig.~\ref{fig:contours}). The colour bar shows the density threshold, $\rho_{\mathrm{thr}}$, of each structure, defined as the minimum density value contained in the structure. \par

For early times, all structures seem to follow a power law relation for both HD and MHD clouds. The exponents were derived from fitting the logarithm of the data points with a non-linear least squares method. For the hydrodynamic clouds we find $n=0.61\pm0.03$, which is slightly steeper than Larson's relation, while the best fit for the MHD clouds gives $n=0.54\pm0.03$ in quite good agreement with the observed scaling relation (Eq.~\ref{eq:larson}), although the scatter is significant.\par
At later times, high density structures tend to trail off to the left of the power law relation, as these sub-structures have decreasing size but retain a high (mass-weighted) velocity dispersion. We will later see that this is a sign of gravity becoming dynamically important. The obvious branch-like collections of points leading towards small scales are actually identified as those dendrogram branches which lead to the densest sub-structures. Due to the presence of these branches, the fitted power law slope is much steeper than the expected $n=0.5$ for later times. From these plots one can also clearly see that the MHD clouds evolve more slowly than the non-magnetized clouds, as the MHD clouds only show dense, flat branches at $t=3.5$~Myr, while these are already present at $t_{\mathrm{evol}}=2.0$~Myr for the HD runs.

\begin{figure*}
    \centering
    \includegraphics[trim=0 0 135 0, clip, height=4.55cm]{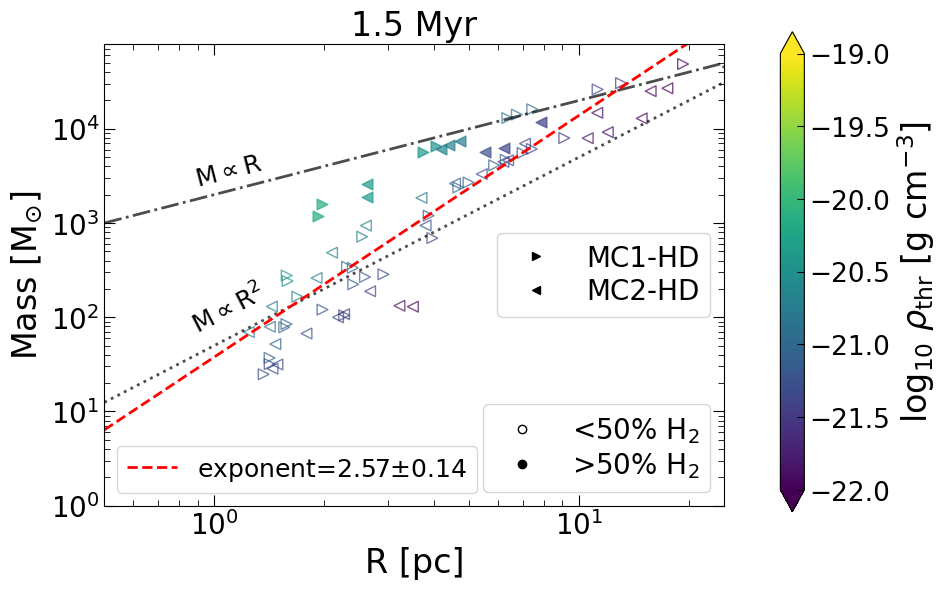}    
    \includegraphics[trim=75 0 135 0, clip, height=4.55cm]{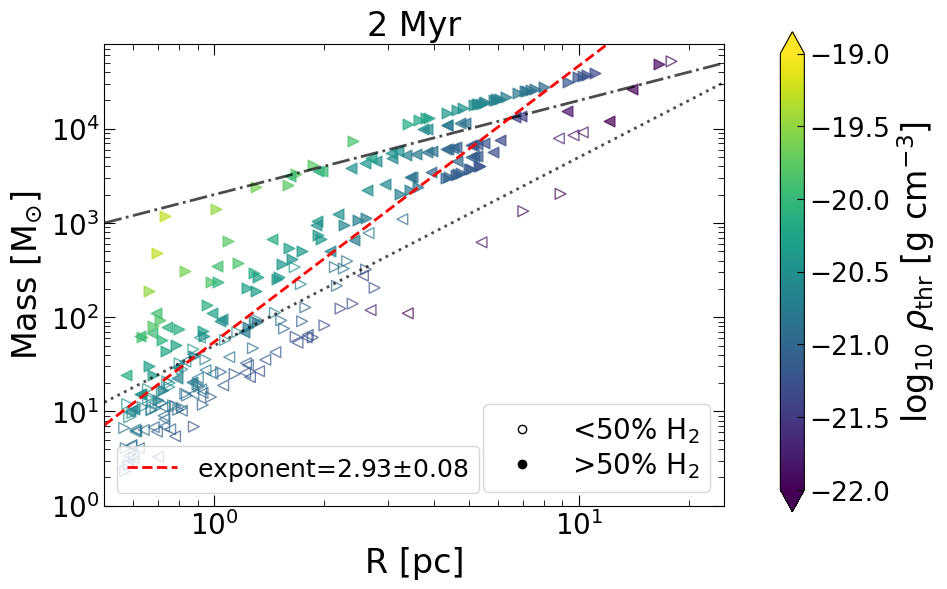}    
    \includegraphics[trim=75 0 0 0, clip, height=4.55cm]{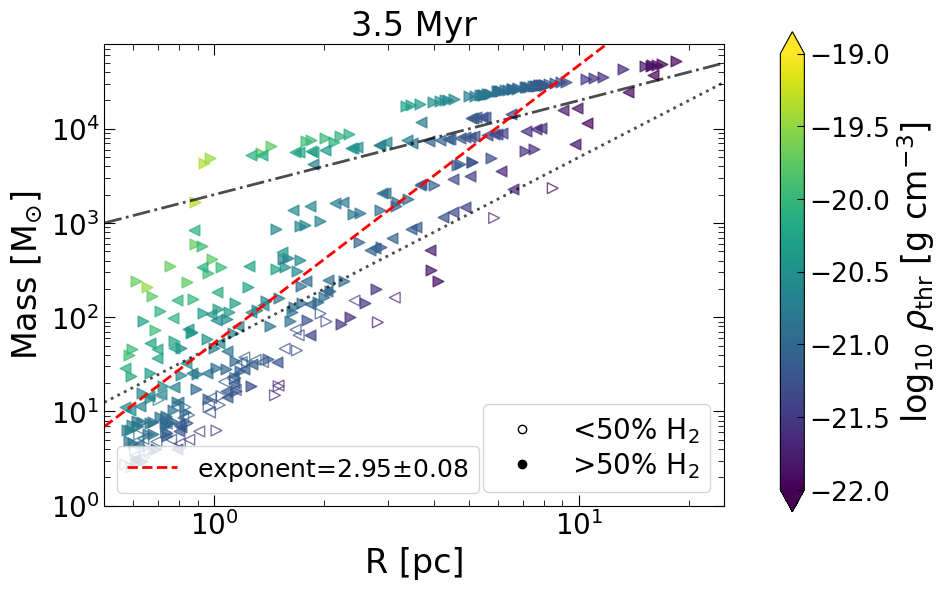}    
    \includegraphics[trim=0 0 135 0, clip, height=4.55cm]{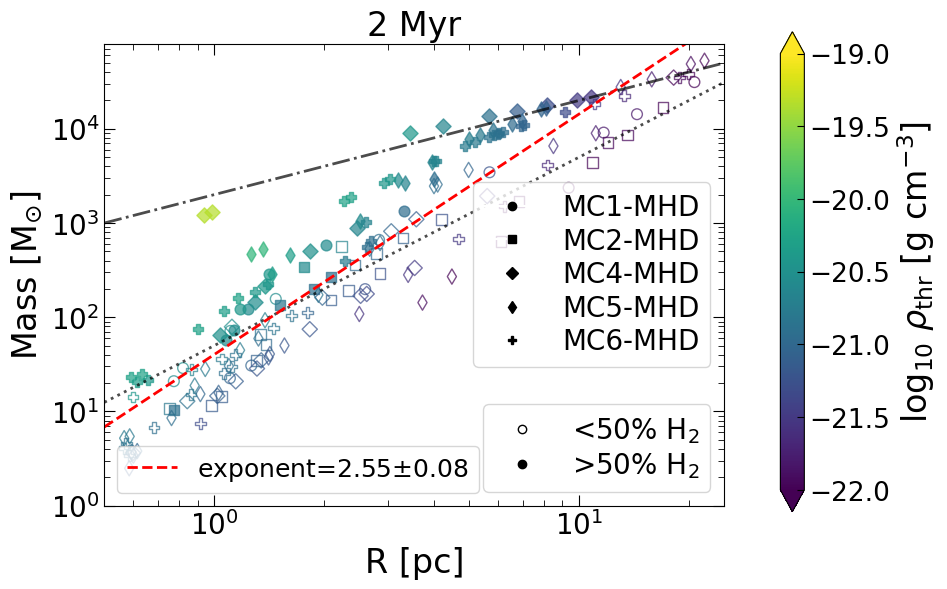}
    \includegraphics[trim=75 0 135 0, clip, height=4.55cm]{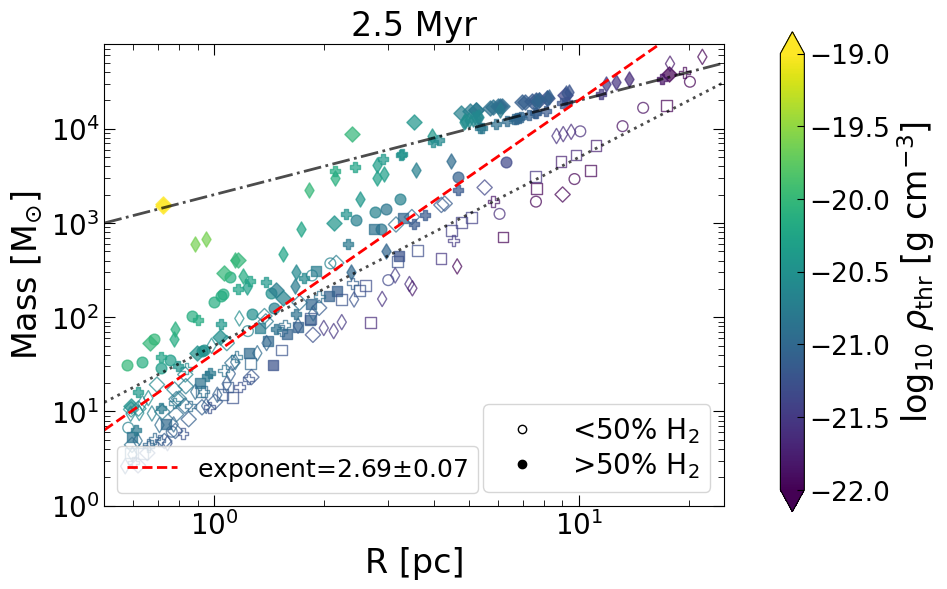}
    \includegraphics[trim=75 0 0 0, clip, height=4.55cm]{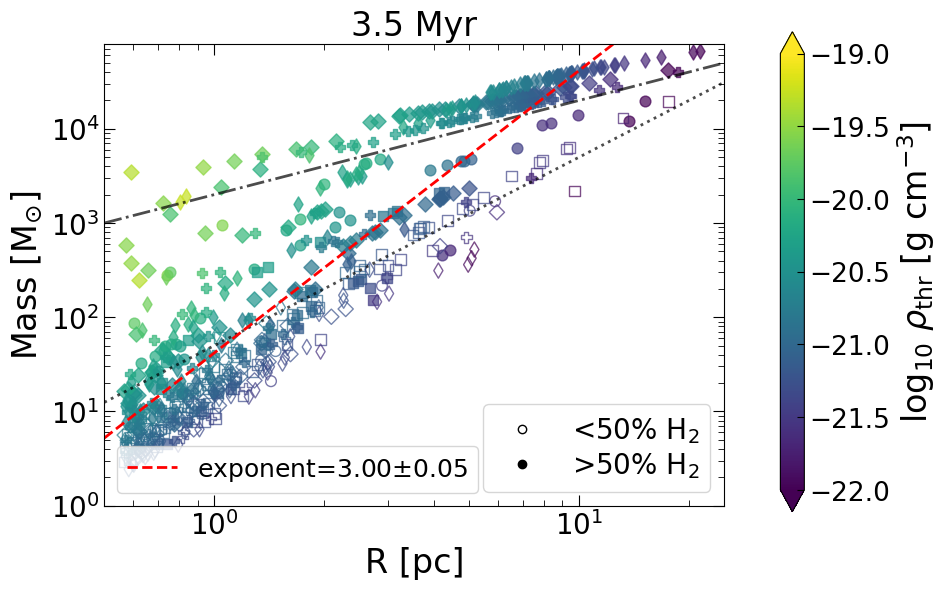}
    \caption{Same as in Fig.~\ref{fig:sigma_size} but now for the mass-size relation. The red, dashed line represents the best-fit, while the black, dash-dotted and the black, dotted lines represent $M\propto R$ and $M\propto R^2$, respectively. The denser branches follow roughly $M\propto R$, while the atomic structures seem to be more consistent with $M\propto R^2$.}
    \label{fig:mass_size}
\end{figure*}

A similar behaviour is seen in the mass-size relation shown in Fig.~\ref{fig:mass_size}. Apart from a best-fit power law, based on a non-linear least squares fit on the logarithm of the values (similar to discussed above), we also plot here lines corresponding to $M\propto R$ and  $M\propto R^2$ in dash-dotted and dotted lines, respectively. The overall fitted slope is generally steeper compared to the $M\propto R^2$ expected from the Larson relation. This steeper slope rather describes the inability of a single exponent to explain the behaviour of all the structures. Visually, the less dense and mostly atomic structures seem to fit better with a $M\propto R^2$ type power law. The denser structures on the other hand clearly branch off to follow a much shallower power law more similar to $M \propto R$. This is reminiscent of more centrally condensed profiles such as Bonnor-Ebert spheres \citep[][]{ebert_uber_1955,bonnor_boyles_1956}, although in our case the structures are far from spherically symmetric. \par 

\begin{figure*}
    \centering
    \includegraphics[trim=0 0 135 0, clip, height=4.55cm]{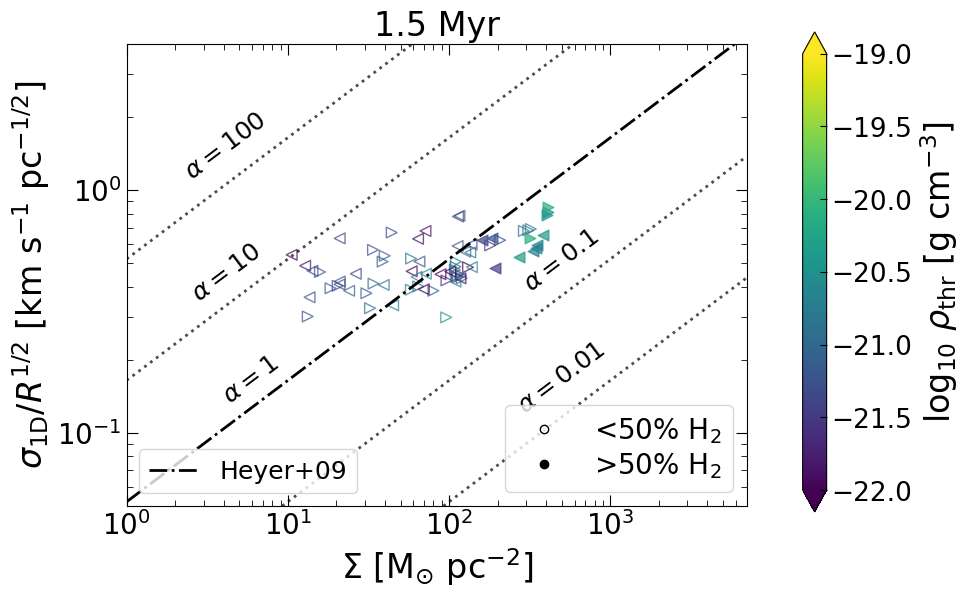}
    \includegraphics[trim=87 0 135 0, clip, height=4.55cm]{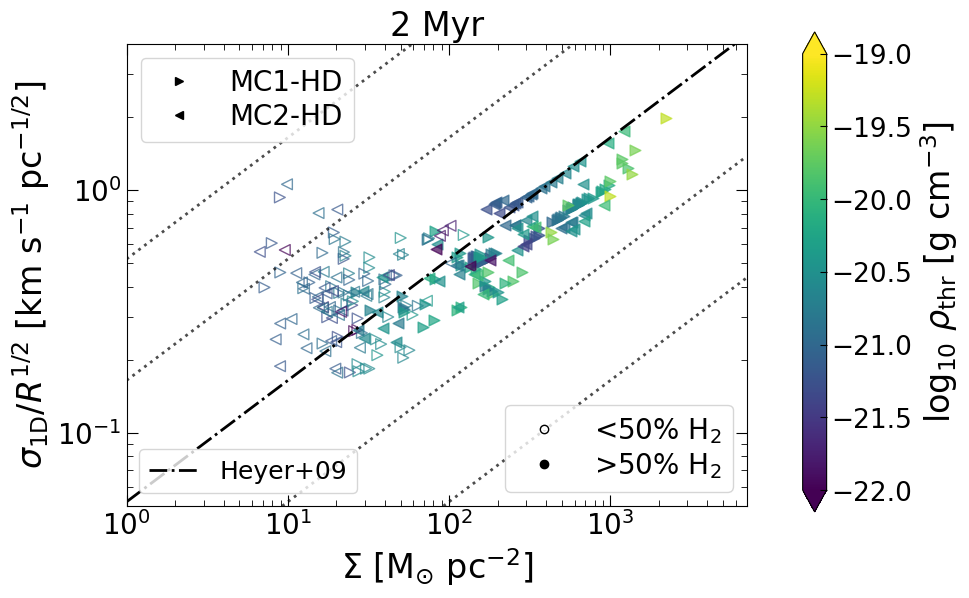}
    \includegraphics[trim=87 0 0 0, clip, height=4.55cm]{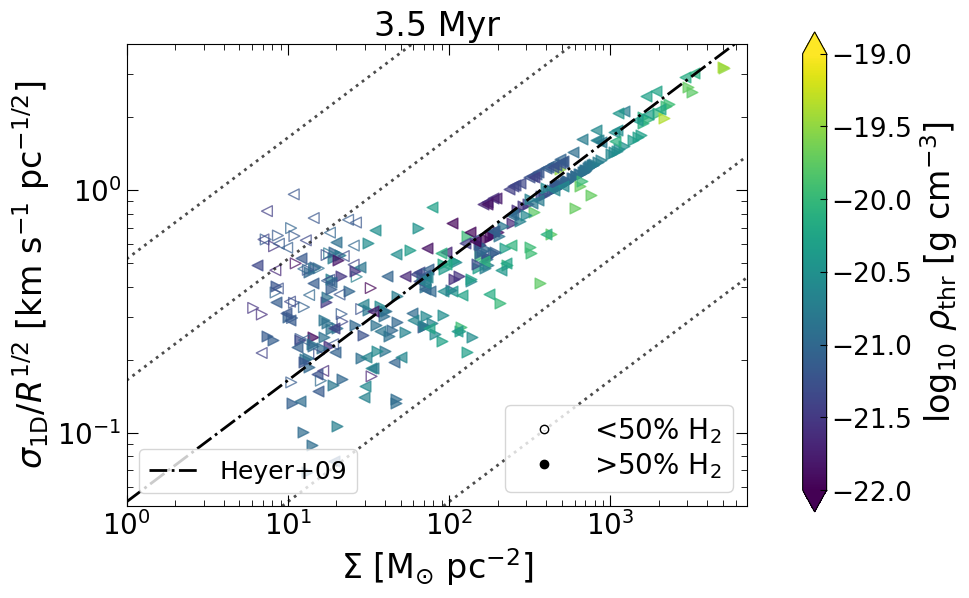}
    \includegraphics[trim=0 0 135 0, clip, height=4.55cm]{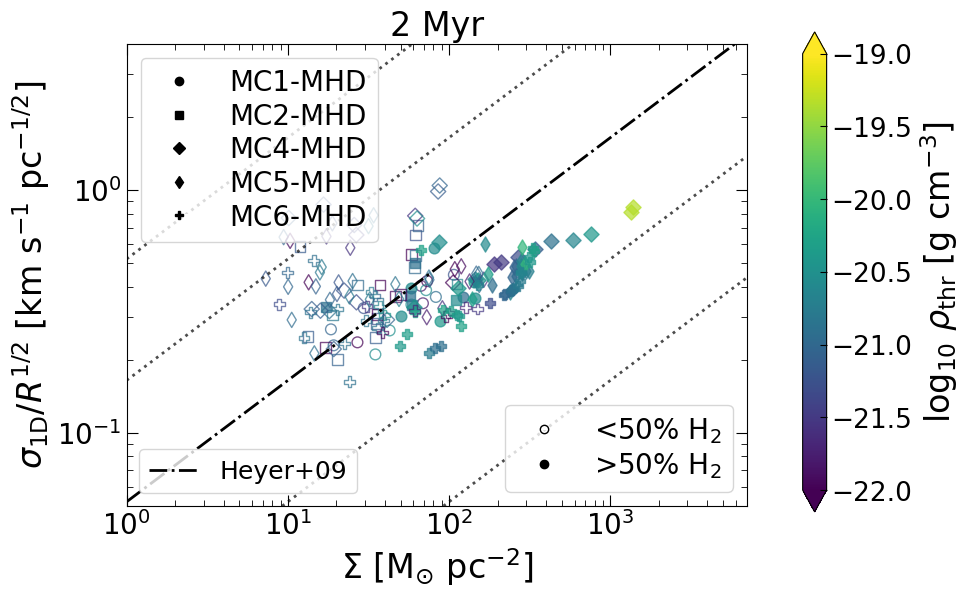}
    \includegraphics[trim=87 0 135 0, clip, height=4.55cm]{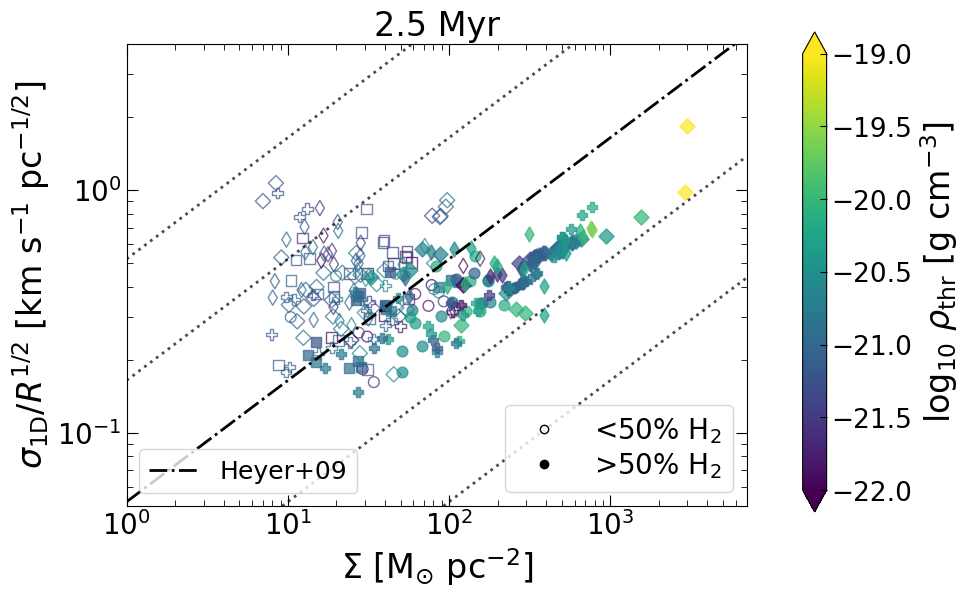}
    \includegraphics[trim=87 0 0 0, clip, height=4.55cm]{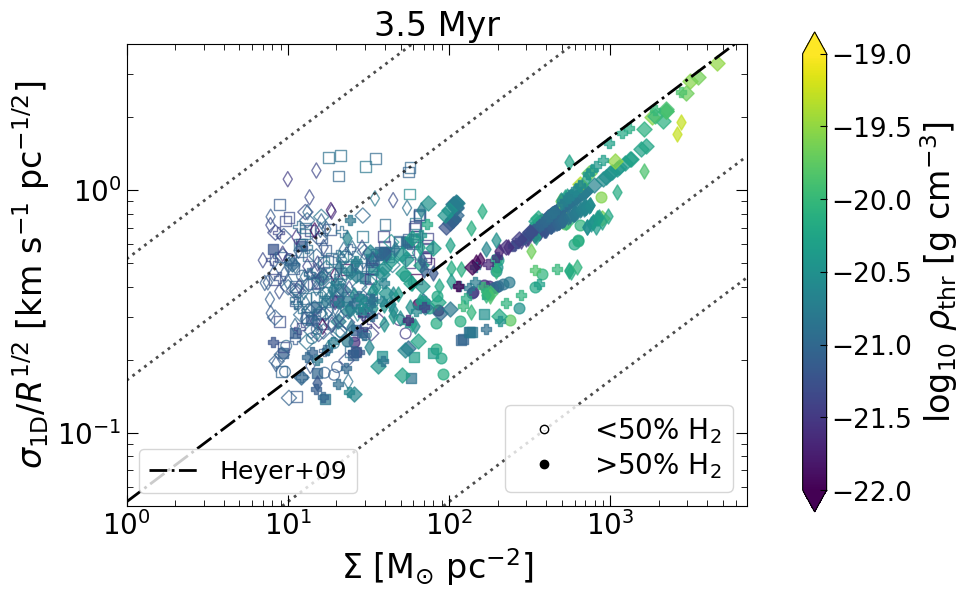}

    \caption{Same as in Fig.~\ref{fig:sigma_size} but now for the Heyer relation. The dash-dotted line represents the $(\pi G \Sigma/5)^{1/2}$ line from \citet{heyer_re-examining_2009} (corresponding to the virial parameter $\alpha=1$), while the parallel dotted lines represent other possible values of $\alpha$ in a purely kinetic virial analysis. We find two populations of structures: denser and mostly molecular structures following the Heyer line towards higher $\Sigma$, and mostly atomic structures that show no correlation with surface density.}
    \label{fig:heyer}
\end{figure*}

Larson's velocity dispersion-size relation was later modified by \citet{heyer_re-examining_2009} to show that the Larson coefficient $\sigma_{\rm 1D}/R^{1/2}$ can depend on the surface density, $\Sigma$, of the structure. We plot the Heyer relation, i.e. $\sigma_{\rm 1D}/R^{1/2}$ vs. $\Sigma$ in figure \ref{fig:heyer}. The surface density of a given structure is calculated as 
\begin{equation}
\Sigma = \rho_{\mathrm{avg}}R.
\end{equation}
Here $\rho_{\mathrm{avg}}$ is the volume-average density of the structure:
\begin{equation}\label{eq:avg_den}
 \rho_{\mathrm{avg}} = M/V.   
\end{equation}
The dash-dotted diagonal line represents the $\sigma_{\rm 1D} = (\pi G/5)^{1/2} \Sigma^{1/2}R^{1/2}$ line from \citet{heyer_re-examining_2009} representing virial equilibrium between kinetic energy and self-gravity ($\alpha=1$, where $\alpha$ is the kinetic virial parameter). Other values of $\alpha$ correspond to lines parallel to the Heyer line in log-log space and are plotted with dotted lines. The virial parameter $\alpha$ obtained in this manner is only a rough representation of the true virial equilibrium, and includes only the kinetic and potential energy terms.  
We find that at earlier times there is very weak, if any, trend along the Heyer line of $\alpha=1$ for both hydro and MHD structures. This picture evolves over time. At $t_{\rm evol}=3.5$~Myr (figure \ref{fig:heyer}, right column), we find two groups of structures: mostly molecular and dense structures that follow the Heyer line, for both HD and MHD clouds and less dense, mostly (but not exclusively) atomic structures that show no correlation at all. \par 
\citet{camacho_energy_2016} have found similar differing behaviour when they look at structures at different density thresholds in turbulent box simulations. Similar results have been also been found by Weis et al., in prep., when analyzing structures forming in colliding flow simulations. We additionally find that the molecular content of the gas is also tracing these two families of structures relatively well. \par 
There are various possible interpretations for structures following the Heyer relation. The original interpretation of \citet{heyer_re-examining_2009} was that they represent gravitational equilibrium. \citet{ballesteros-paredes_gravity_2011} point out that, given the typical error estimates, they could also be consistent with gravitational infall, as the infall velocity and virial equilibrium velocity only differ by a factor of $\sqrt{2}$. A similar discussion was also presented in \citet{larson_turbulence_1981}. In addition, the estimate of the surface density here is rather crude, and does not incorporate the fractal nature of the complicated medium. To investigate different possible scenarios, we perform an analysis on the different energy terms in the next section. \par

\section{energetics}\label{sec:energetics}

The interplay of various forces determines the dynamics of the structures involved. For a structure embedded in a more diffuse medium, we consider the self-gravity, kinetic energy, as well as thermal and magnetic energy (for the MHD runs) of the structure. We limit ourselves to the discussion of the various volume terms and do not consider the time dependent or the surface terms of the full virial theorem. Below, we briefly discuss how we determine the various terms.  \par
\subsection{Energy terms}
The gravitational energy of a structure can be separated in two parts - the self-gravity due to its own mass, and the gravitational energy due to the fact that the structure is embedded in an asymmetric gravitational potential caused by the surrounding gas. For the analysis of this section, we have limited ourselves to considering only self-gravity for a more traditional virial analysis.  In section \ref{sec:gravity}, we analyse the relative importance of the external medium.\par
The self gravitational energy of the structure is defined as 
\begin{equation}
E_{\mathrm{PE}} = -\frac{1}{2}G\int_V \int_V \frac{\rho(\mathbf{r}) \rho(\mathbf{r}')}{|\mathbf{r} -\mathbf{r}'|} \mathrm{d}^3r \mathrm{d}^3r' .
\end{equation}
We compute this term by using a KD tree with a geometric opening angle of 0.5 radians. By comparing with the exact potential computed from a direct sum, we find that the errors introduced by the tree method are typically $\sim$ 1\%. \par
We also compute the volume term of the magnetic energy, which can contribute to the important magnetic pressure term. This pressure term can act against compression and prevent a structure from collapsing. The magnetic energy of a given structure is computed as 
\begin{equation}
E_{\mathrm{B}} = \frac{1}{8\pi}\int_V |\mathbf{B}|^2 \mathrm{d}^3r ,
\end{equation}
where $|\mathbf{B}|$ is the magnitude of the magnetic field.\par
We compute the thermal energy of a structure as 
\begin{equation}
E_{\mathrm{TE}} = \frac{3}{2}\int_V P \mathrm{d}^3r,
\end{equation}
where $P$ is the thermal pressure of the gas.\par
The computation of the kinetic energy, $E_{\mathrm{KE}}$, of a structure is done as
\begin{equation}
    E_{\mathrm{KE}} = \frac{1}{2}\int_V\rho(\mathbf{v}-\mathbf{v}_0)^2\mathrm{d}^3r,
\end{equation}\label{eq:ke}
where $\mathbf{v}_0$ is the bulk velocity computed from Eq.~\ref{eq:v0}.\par 
From the above quantities, we estimate a virial ratio of the volume energy terms:
\begin{equation}\label{eq:virial}
    \alpha_{\mathrm{vir}}^{\mathrm{vol}} = \frac{2E_{\mathrm{KE}}+2E_{\mathrm{TE}}+E_{\mathrm{B}}}{|E_{\mathrm{PE}}|}.
\end{equation}
The magnetic energy is absent for purely hydrodynamic simulations. Eq. \ref{eq:virial} is a simplification of the full virial theorem \citep[][]{mckee_virial_1992, dib_virial_2007}, which includes both time-dependent as well as surface terms, which is beyond the scope of this paper. However, we discuss the surface terms briefly at the end of Section \ref{sec:virial}.\par
We define bound structures as those with a virial ratio $\alpha_{\mathrm{vir}}^{\mathrm{vol}}<1$ and marginally bound structures as $1\leq\alpha_{\mathrm{vir}}^{\mathrm{vol}}<2$. The definition of marginally bound structures here is motivated by the fact that, in presence of only self-gravity and kinetic energy, a structure with $\alpha_{\mathrm{vir}}^{\mathrm{vol}}=2$ would have a net zero total energy, $E_{\rm tot}$, i.e.
\begin{equation}
    E_{\rm tot} = E_{\rm KE} + E_{\rm PE} = 0,
\end{equation}
and as a consequence,
\begin{equation}
    \alpha = \frac{2E_{\rm KE}}{|E_{\rm PE}|},
\end{equation}
is reduced to $\alpha=2$.
\subsection{Virial analysis}\label{sec:virial}
\begin{figure*}
    \centering
    \includegraphics[trim=0 30 0 5, clip, width=0.49\textwidth]{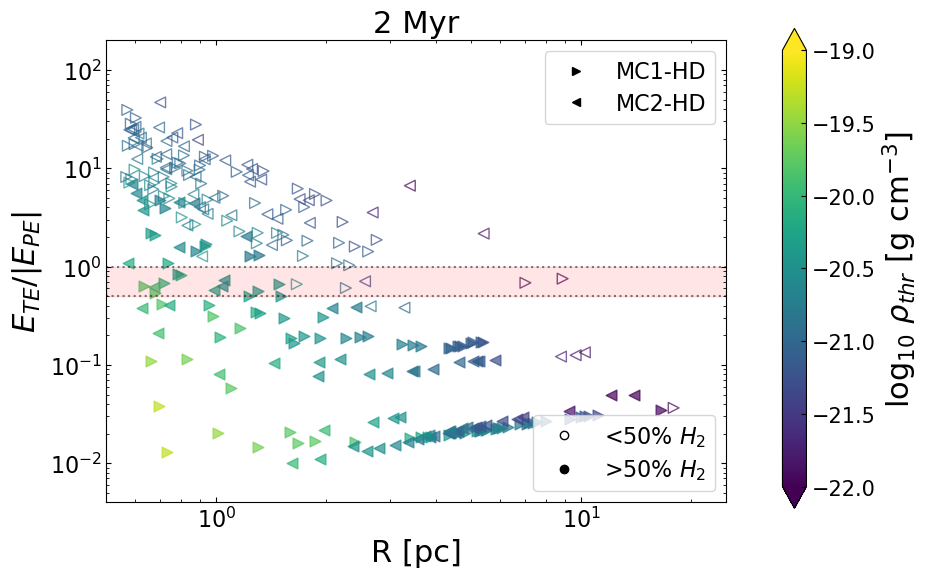}
    \includegraphics[trim=0 30 0 5, clip, width=0.49\textwidth]{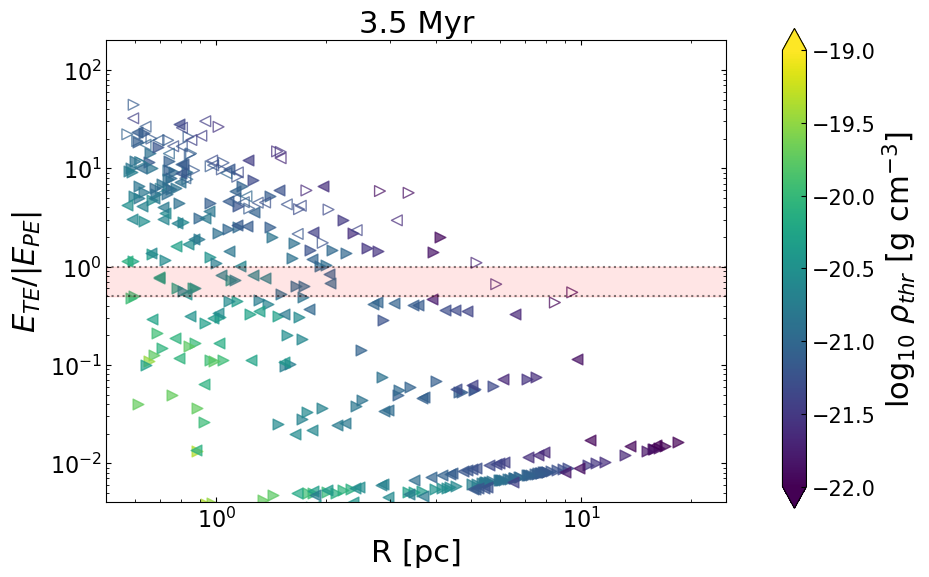}

    \includegraphics[trim=0 30 0 5, clip, width=0.49\textwidth]{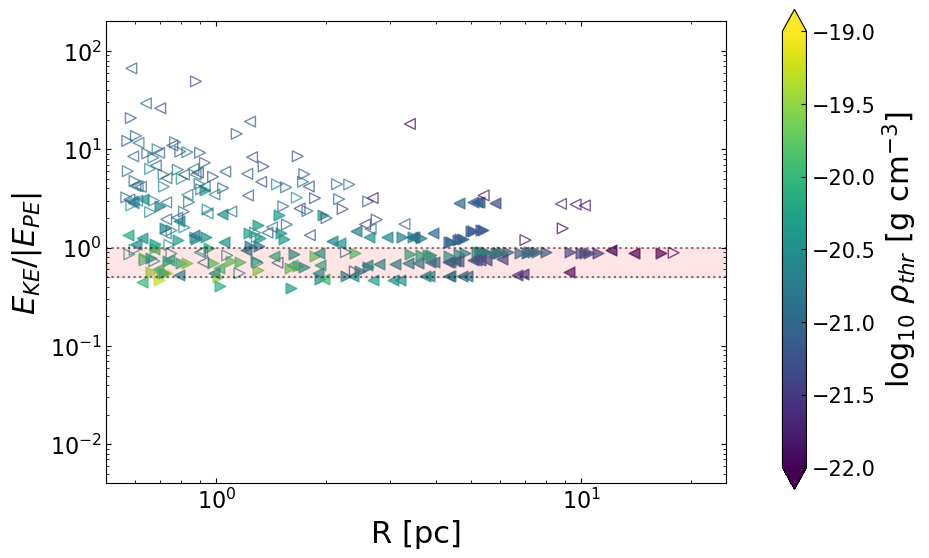}
    \includegraphics[trim=0 30 0 5, clip, width=0.49\textwidth]{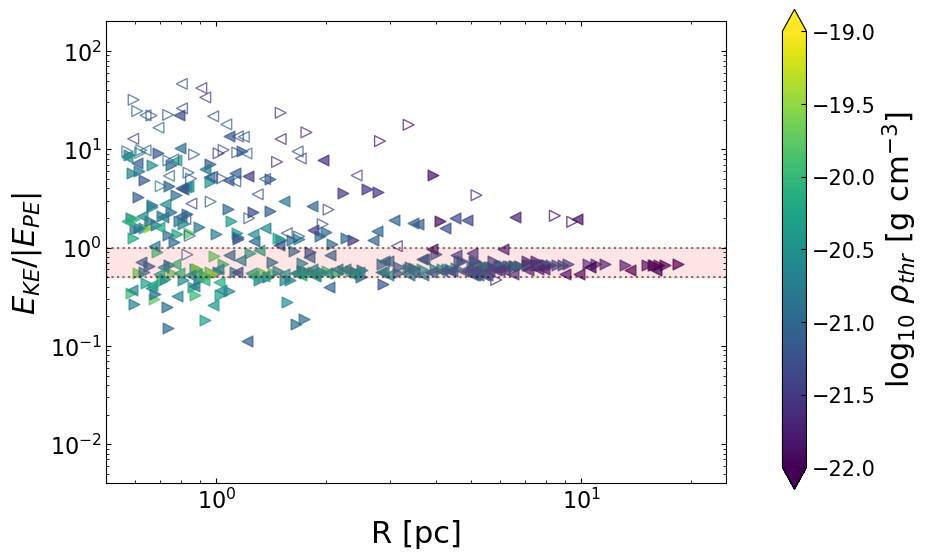}

    \includegraphics[trim=0 8 0 5, clip, width=0.49\textwidth]{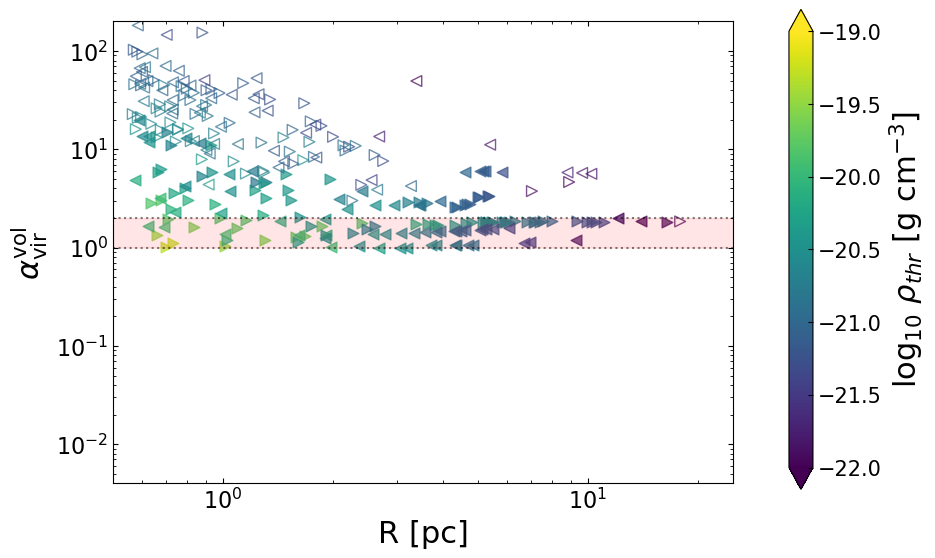}
    \includegraphics[trim=0 8 0 5, clip, width=0.49\textwidth]{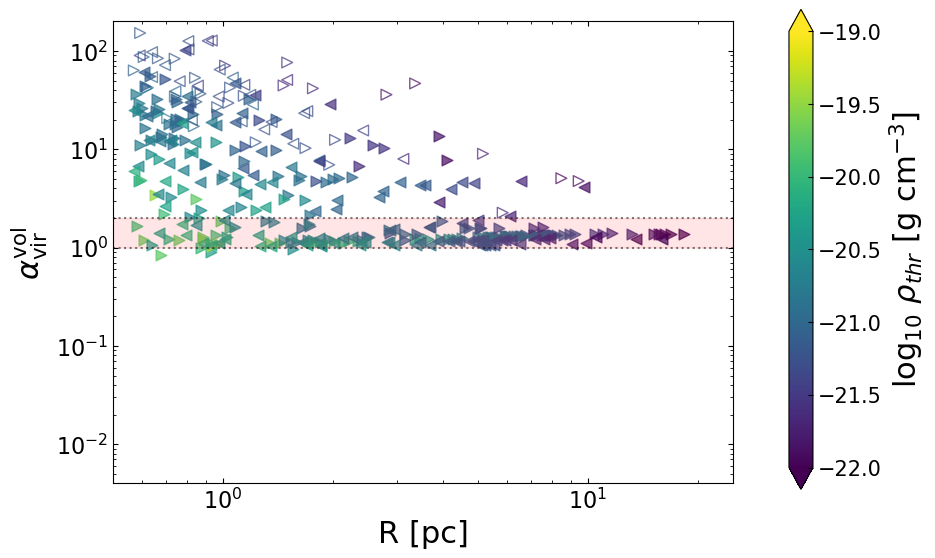}

    \caption{Energetics of the two HD clouds at $t_{\rm evol}=2$~Myr (left column) and $t_{\rm evol}=3.5$~Myr (right column). Each plot is a ratio of a different energy to the potential energy. The red shaded region represents the region between marginally bound and virialized ($1\leq \alpha<2$) for that particular energy term. 
    Large scale structures are marginally bound in the virial analysis. At smaller scales, most of the structures are kinetically dominated with only a few marginally bound structures, as well as a small number of bound structures.}
    \label{fig:HD_energy}
\end{figure*}

\begin{figure*}
    \centering
    \includegraphics[trim=0 30 0 5, clip, width=0.49\textwidth]{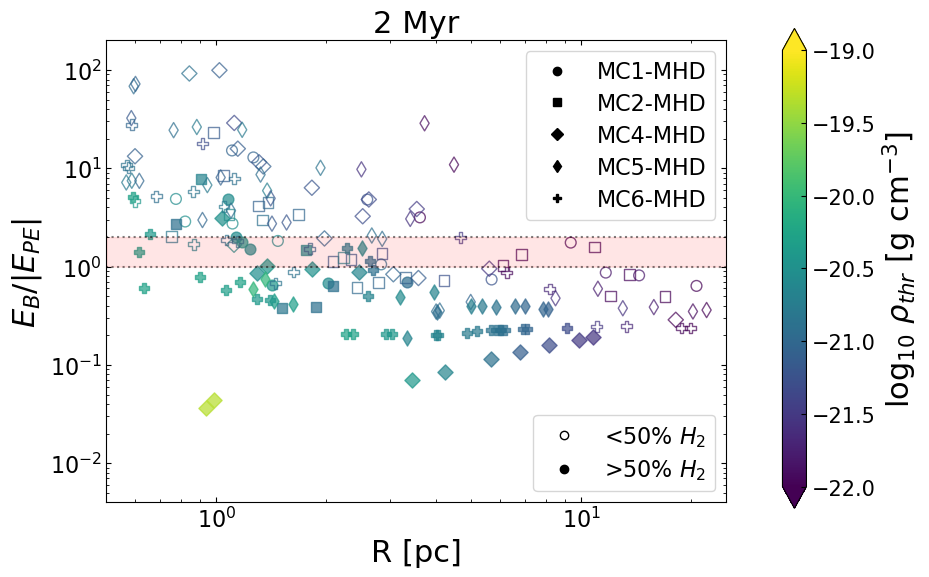}
    \includegraphics[trim=0 30 0 5, clip, width=0.49\textwidth]{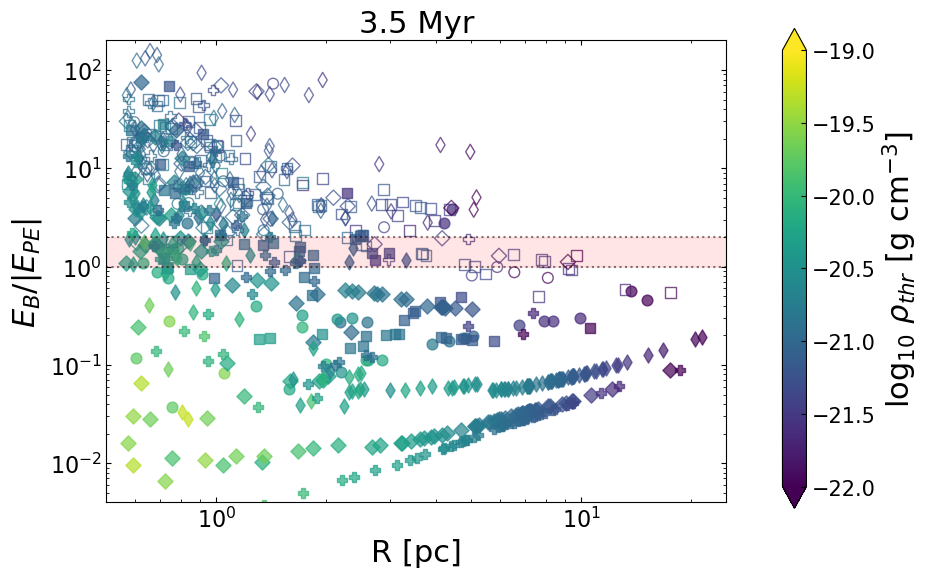}

    \includegraphics[trim=0 30 0 5, clip,width=0.49\textwidth]{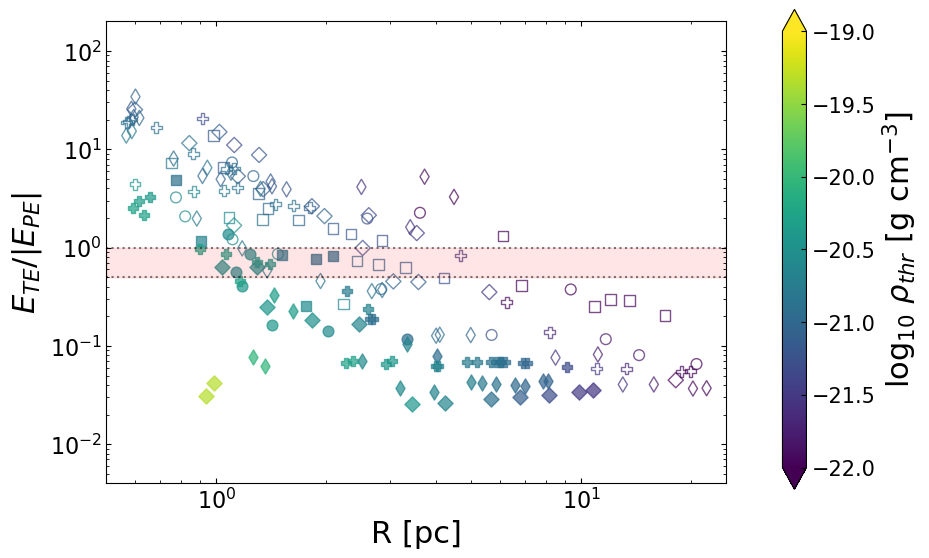}
    \includegraphics[trim=0 30 0 5, clip,width=0.49\textwidth]{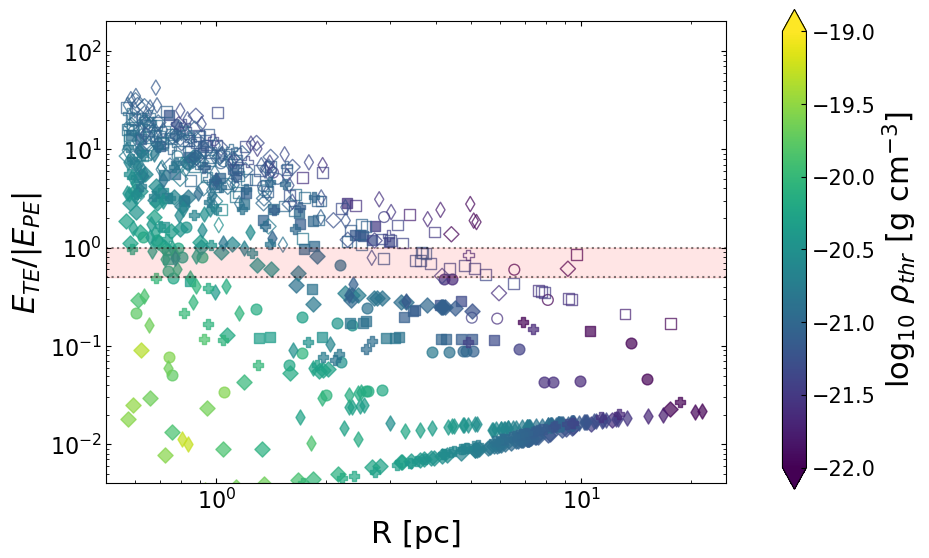}

    \includegraphics[trim=0 30 0 5, clip,width=0.49\textwidth]{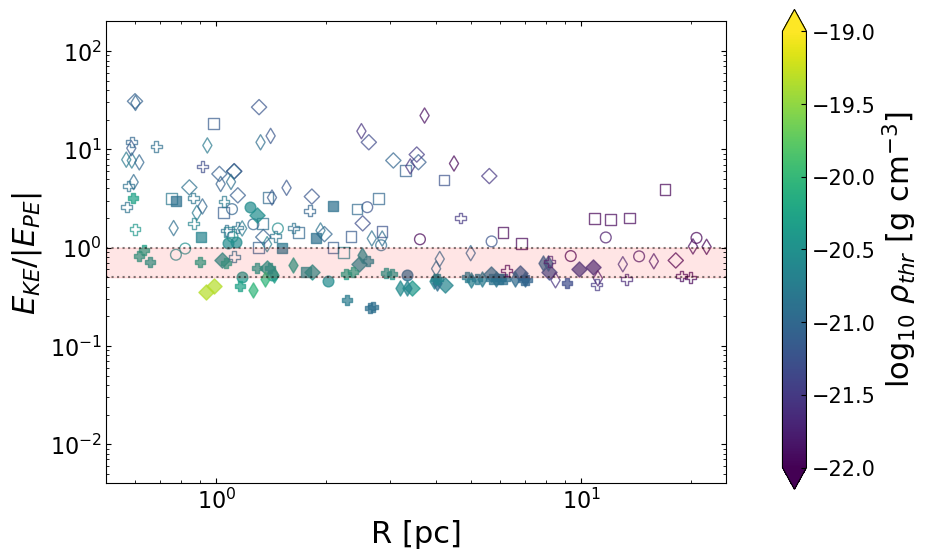}
    \includegraphics[trim=0 30 0 5, clip,width=0.49\textwidth]{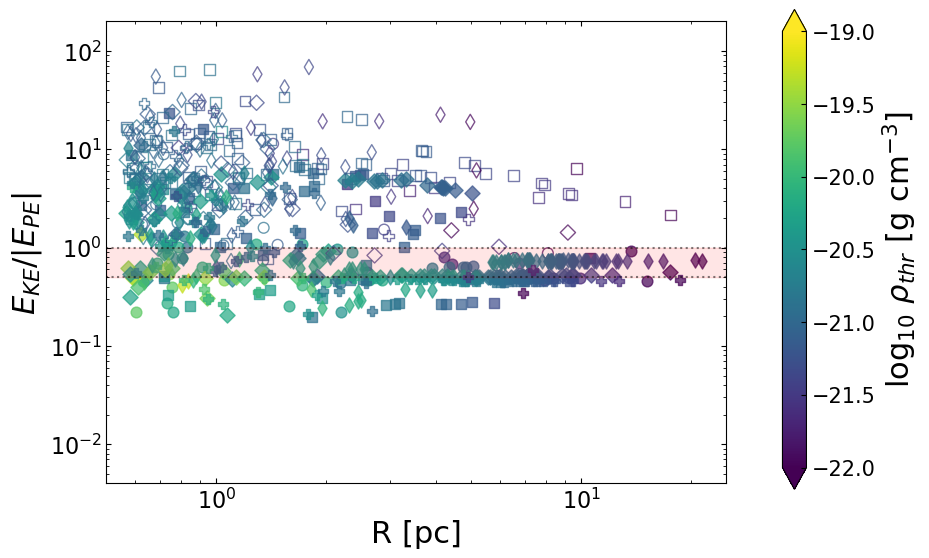}

    \includegraphics[trim=0 8 0 5, clip, width=0.49\textwidth]{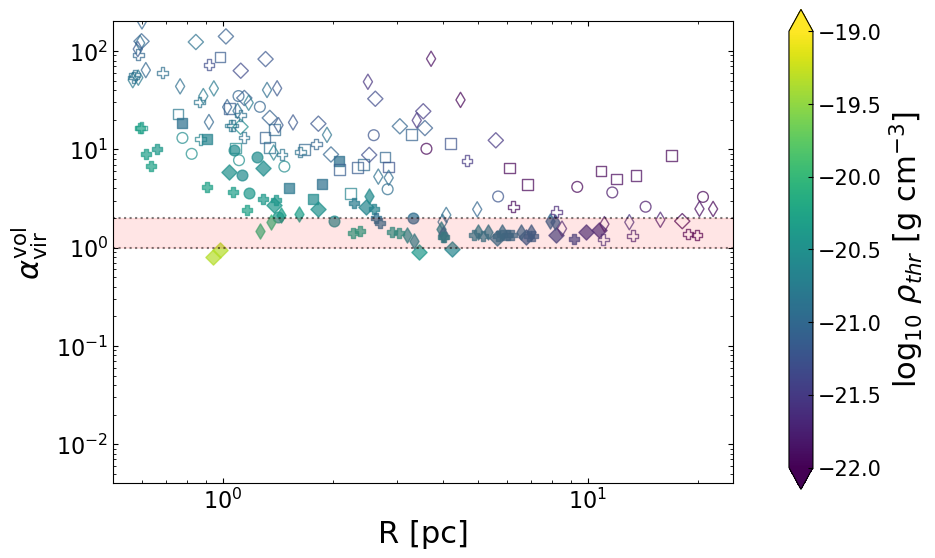}
    \includegraphics[trim=0 8 0 5, clip, width=0.49\textwidth]{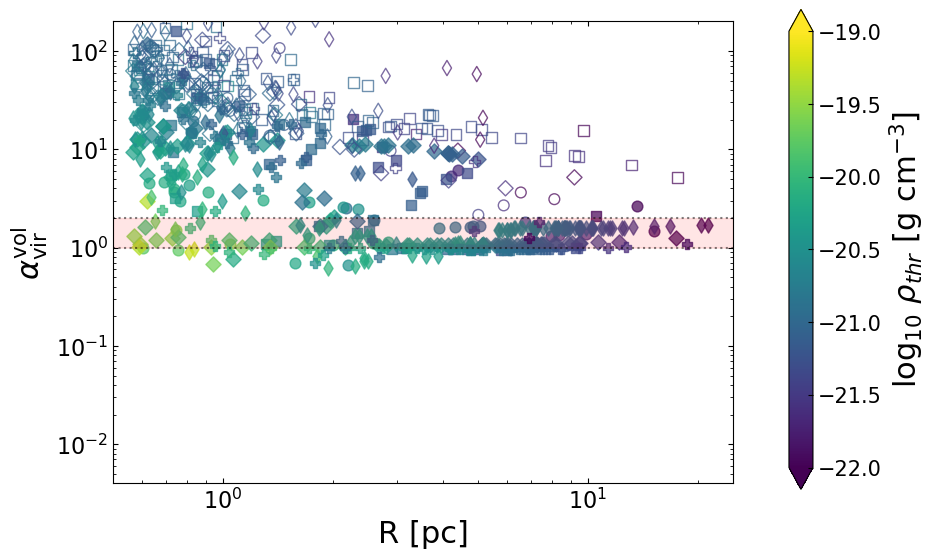}

    \caption{Same as in Fig~\ref{fig:HD_energy} but now for the MHD clouds. Additionally $E_{\rm B}/|E_{\rm PE}|$ is plotted in the top row. The red shaded region represents the region between marginally bound and virialized ($1\leq \alpha<2$) for that particular energy term.
    Note that due to Eq. \ref{eq:virial}, this implies a factor of 2 difference between the energy ratios and $\alpha_{\mathrm{vir}}^{\mathrm{vol}}$, except for $E_{\mathrm{B}}/|E_{\mathrm{PE}}|$. Large scale structures are mostly marginally bound in the virial analysis. Gravitationally bound structures are few and emerge over time, suggesting their origin is from the dynamics of the marginally bound gas.}
    \label{fig:MHD_energy}
\end{figure*}

In Fig. \ref{fig:HD_energy}, we show the time evolution of the ratio of different energy terms and the self gravitational energy, as well as $\alpha_{\rm vir}^{\rm vol}$ (bottom row) as a function of their size $R$ (see Eq.~\ref{eq:size})  for the two hydrodynamic clouds and in Fig. \ref{fig:MHD_energy} for the different MHD clouds. We plot the sub-structures at 2 Myr (left column) and at 3.5 Myr (right column). For the HD clouds (Fig. \ref{fig:HD_energy}), the top two rows present the evolution of thermal and kinetic energy over $E_{\rm PE}$ as a function of size $R$. The colors represent the density threshold of each analyzed structure. For the MHD clouds (Fig. \ref{fig:MHD_energy}), we also plot the magnetic energy over $E_{\rm PE}$ in the top row. The red shaded region shows the marginally bound region. According to Eq. \ref{eq:virial}, for the ratio of magnetic to potential energy, marginally bound is where $1\leq E_{\mathrm{B}}/|E_{\mathrm{PE}}|<2$; for the kinetic and thermal energies, this represents a region where $0.5\leq E_{\mathrm{KE}}/|E_{\mathrm{PE}}|<1$ and $0.5\leq E_{\mathrm{TE}}/|E_{\mathrm{PE}}|<1$, respectively. This results $1\leq \alpha_{\rm vir}^{\rm vol}<2$.\par

For the sub-structures in the HD clouds (Fig. \ref{fig:HD_energy}), we find that the thermal energy is mostly irrelevant for the large scale, as well as for the denser and molecular structures. However, it seems to be important for atomic structures on all scales and for some small-scale molecular structures, particularly at later times. The $E_{\rm KE}/|E_{\rm PE}|$ ratio is far closer to unity for large scale ($R$>10 pc) and denser ($\rho_{\rm thr}\gtrapprox 10^{-20}$~g~cm$^{-3}$) structures. Due to the significant contribution of $E_{\rm KE}$, we find that only a few structures seem to be bound, although the number of such structures increases from 2 to 3.5 Myr. When we consider the full virial parameter, the behaviour is quite similar to that of the kinetic energy, with the addition that even fewer structures are bound at all times. The emergence of gravitationally bound structures over time suggests a growing importance of gravity as the cloud becomes more evolved. This complements the picture we see in Larson's $\sigma_{\rm 1D}$-$R$ relation (Fig. \ref{fig:sigma_size}), where a flat line-width size relation emerges for the densest branches over time. \par
For the MHD clouds (Fig. \ref{fig:MHD_energy}), the magnetic energy is relevant only for less dense and mostly atomic structures, and does not appear to be significant for the larger scale structures or the denser and potentially star forming structures. This is true at earlier as well as later times. This behaviour is echoed in the evolution of $E_{\rm TE}/|E_{\rm PE}|$, which is quite similar to the behaviour in the hydrodynamic clouds. As in the hydrodynamic case, the dynamics of larger scale, as well as dense structures is mostly dominated by the interplay of $E_{\rm KE}$ and $E_{\rm PE}$. The behaviour of $\alpha_{\mathrm{vir}}^{\mathrm{vol}}$ for HD and MHD clouds is also similar and follows the overall trend of $E_{\rm KE}/|E_{\rm PE}|$. We note one exception to the purely hydrodynamic structures here though. While at earlier times, we barely have any bound structures at 2 Myr, this number is much higher at 3.5 Myr. This suggests that, with magnetic fields, bound structures form more slowly from a marginally bound environment.\par
The emergence of gravitationally bound structures over time from a marginally bound environment, suggests that gravity only gradually becomes important for the densest branches, but is not the primary driving force for the formation of the structures. Our findings are more in line with the GT scenario of structure formation, where gravity takes over only once the turbulent structures are sufficiently compressed.\par
We stress that the aspect of gradually increasing boundness is essential here.  If we look at only the relatively late stages at $t_{\rm evol}=3.5$~Myr for the large-scale cloud structures, which are all molecular and lie close to $\alpha_{\rm vir}^{\rm vol}=1$, it is difficult to distinguish between GT and GHC. This can be resolved only by looking at earlier times. The behaviour of the kinetic virial ratio, as well as $\alpha_{\rm vir}^{\rm vol}$, for the largest-scale structure in each cloud can be seen in Table \ref{tab:alpha_large_scale}. We find that at 2 Myr, almost all largest-scale structures, with the exception of MC6-MHD, have $\alpha_{\rm vir}^{\rm vol}>1.8$, with three of the clouds having $\alpha_{\rm vir}^{\rm vol}>2$. This is dramatically reduced at later times, with only MC2-MHD being unbound ($\alpha_{\rm vir}^{\rm vol}>2$) at $t_{\rm evol}=3.5$~Myr.\par 
\begin{table}
    \centering
    \begin{tabular}{@{\extracolsep{\fill}}*{6}{c}}
    \hline
        Run & \multicolumn{2}{c}{2 Myr} && \multicolumn{2}{c}{3.5 Myr}\\
        \cline{2-3} \cline{5-6}
         name& $2E_{\rm KE}/|E_{\rm PE}|$ & $\alpha_{\rm vir}^{\rm vol}$&&$2E_{\rm KE}/|E_{\rm PE}|$&$\alpha_{\rm vir}^{\rm vol}$\\
         \hline
    MC1-HD & 1.77 & 1.85 &&1.35&1.39\\
    MC2-HD & 1.75 & 1.84 &&1.19&1.21\\
    \hline 
    MC1-MHD&2.51&3.28&&0.93&1.47\\
    MC2-MHD&7.81&8.71&&4.30&5.18\\
    MC4-MHD&1.48&1.86&&1.13&1.27\\
    MC5-MHD&2.04&2.47&&1.48&1.71\\
    MC6-MHD&1.00&1.35&&0.95&1.09\\
    \hline
    \end{tabular}
    \caption{The kinetic virial ratio and $\alpha_{\rm vir}^{\rm vol}$ for the largest-scale ($\sim 20$~pc) dendrogram structures for different HD and MHD clouds at 2 and 3.5 Myr. All large scale structures have $\alpha_{\rm vir}^{\rm vol}>1$, but the value reduces considerably from 2 to 3.5 Myr.}
    \label{tab:alpha_large_scale}
\end{table}
The decrease in $\alpha_{\rm vir}^{\rm vol}$ for large scale structures which are unbound or marginally bound suggests that the clouds are being compressed, likely by larger scale flows, towards becoming gravitationally bound (see also Section \ref{sec:infall}). While we explicitly show this here only for the largest structures, the dependence of $\alpha_{\rm vir}^{\rm vol}$ on $R$ is relatively flat at large scales ($R>10$~pc, Figs. \ref{fig:HD_energy} and \ref{fig:MHD_energy}, bottom panels) and we can therefore conclude that this trend is not a consequence of e.g. the choice of our initial dendrogram threshold. \par
\begin{figure*}
    \centering
    \includegraphics[width=0.49\textwidth]{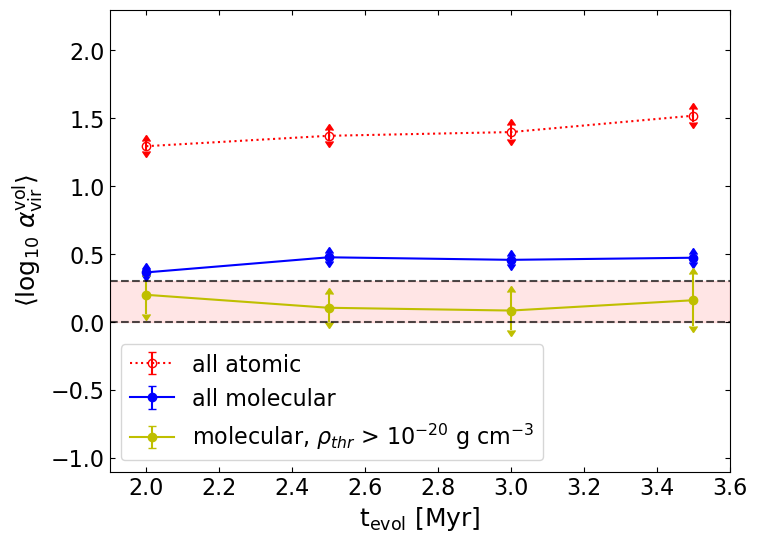}
    \includegraphics[width=0.49\textwidth]{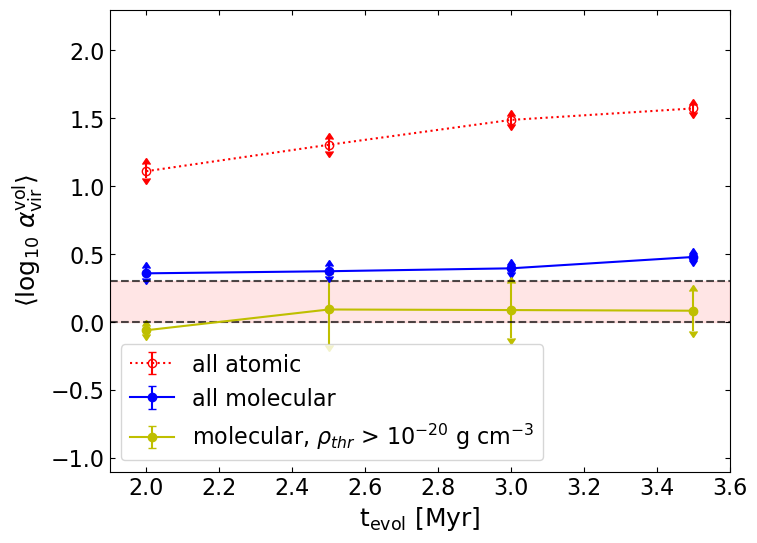}
    \caption{Time evolution of the average behaviour of $\alpha_{\mathrm{vir}}^{\mathrm{vol}}$ for HD (left) and MHD clouds (right). The points represent the mean of log$_{10}$~$\alpha_{\mathrm{vir}}^{\mathrm{vol}}$, while the error bars represent the error on the mean. There is a clear distinction between the average behaviour of these ratios between atomic structures on the one side, and molecular as well as dense structures on the other side. This suggests that the molecular content helps us distinguish between two kinds of structures - unbound and mostly atomic, and marginally bound and mostly molecular.}
    \label{fig:alpha_time_evol}
\end{figure*}

While over time we do see the emergence of individual structures below the bound line, as well as a decrease in the value of $\alpha_{\rm vir}^{\rm vol}$ at large scales, the overall statistical behaviour of most of the cloud sub-structures remains more or less unchanged. This can be seen in Fig. \ref{fig:alpha_time_evol}, where we show the time evolution of the average of the logarithm of $\alpha_{\mathrm{vir}}^{\mathrm{vol}}$, $\langle {\rm log}_{10}\ \alpha_{\mathrm{vir}}^{\mathrm{vol}} \rangle$, for atomic, molecular, and dense molecular structures (structures which are molecular and have a threshold density $\rho_{\rm thr}>10^{-20}$~g~cm$^{-3}$), respectively. Here, we sum over all the different cloud structures, differentiating only between hydrodynamic (Fig. \ref{fig:alpha_time_evol}, left panel) and MHD clouds (Fig. \ref{fig:alpha_time_evol}, right panel). The vertical error bar shows the standard error on the mean of log$_{10}\ \alpha_{\mathrm{vir}}^{\mathrm{vol}}$. Fig.~\ref{fig:alpha_time_evol} illustrates that the atomic and molecular structures have clearly different energetic behaviour. We can attribute this to the fact that they trace different parts of the parent cloud, as can be visually seen in Fig. \ref{fig:contours}. The chemical evolution followed in the simulations seems to trace the dynamical difference of the cloud sub-structures in different environments. We further observe that, as we trace denser gas, the average value of $\alpha_{\rm vir}^{\rm vol}$ decreases, suggesting that we trace gas which is close to being gravitationally bound. \par

As a conclusion to the virial analysis, we briefly remark on the expected behaviour of the surface terms which also contribute to the full virial ratio. Since we find that for most of the structures whose dynamics is important for eventual star formation (both large scale, as well as denser), the kinetic and the potential energy dominate their dynamics, we expect that their corresponding surface terms could be important as well. We perform a detailed anaylsis on the tidal effect of the external medium in Section \ref{sec:gravity}. 
\subsection{Inflow or infall}\label{sec:infall}

\begin{figure*}
    \centering
    \includegraphics[trim=0 30 0 5, clip,width=0.49\textwidth]{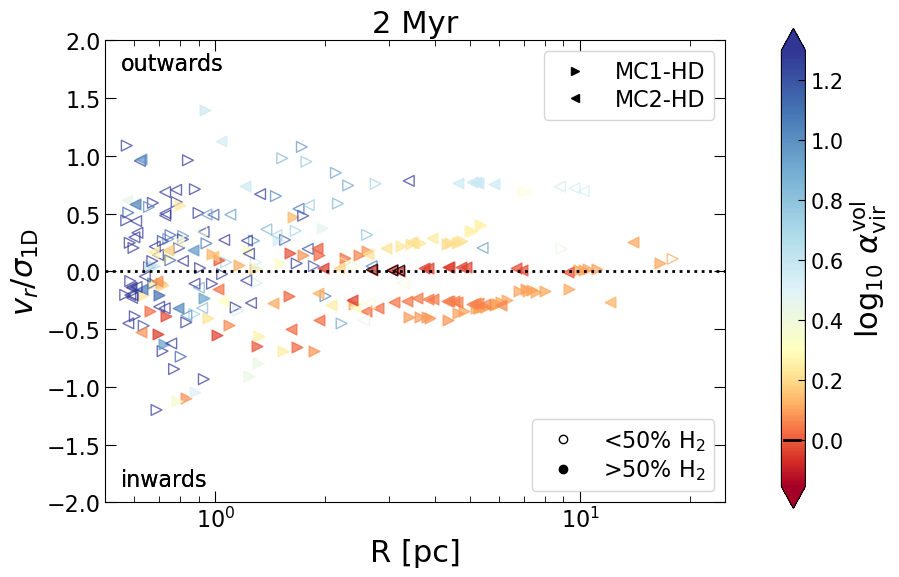}
    \includegraphics[trim=0 30 0 5, clip,width=0.49\textwidth]{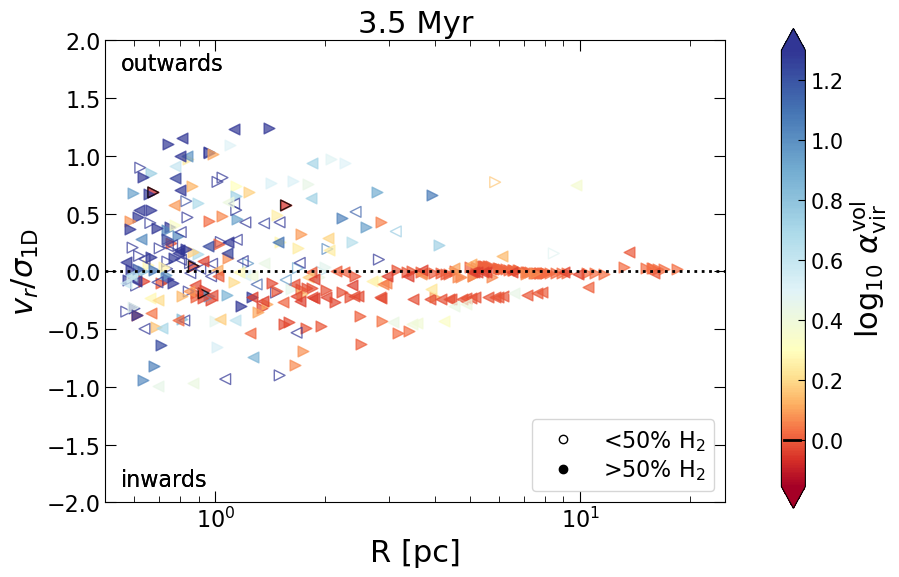}
    \includegraphics[trim=0 8 0 5, clip,width=0.49\textwidth]{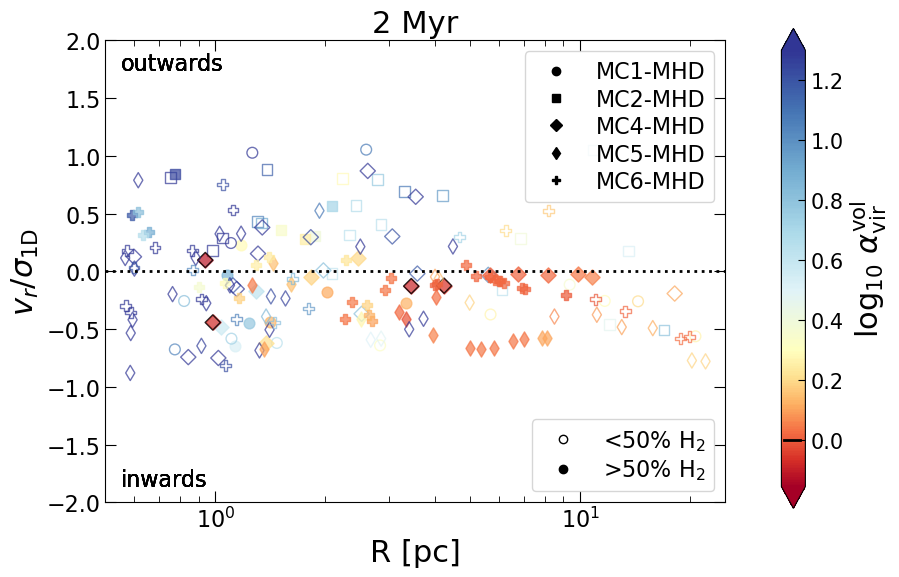}
    \includegraphics[trim=0 8 0 5, clip,width=0.49\textwidth]{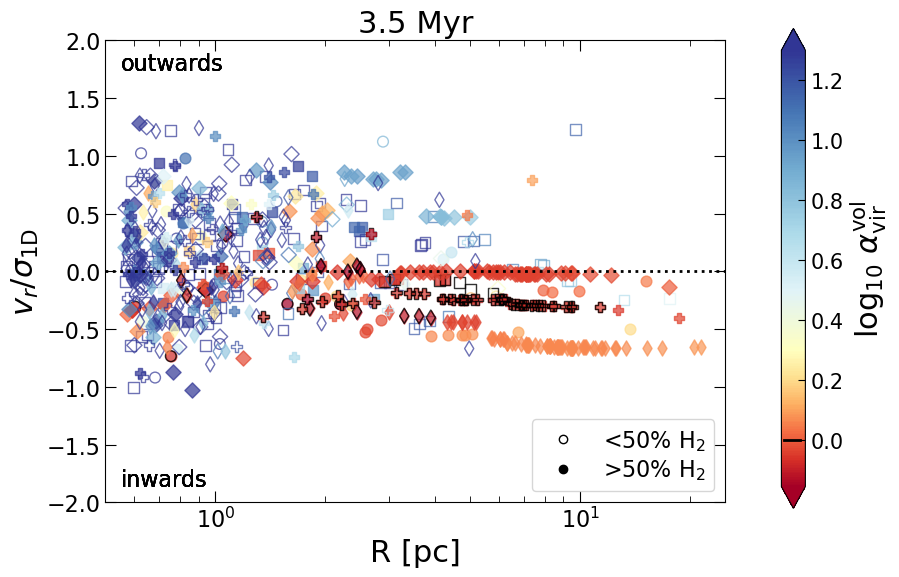}
    \caption{Radial velocity over 1D velocity dispersion plotted against the size for HD clouds at $t_{\rm evol}$=2 Myr (top left) and $t_{\rm evol}$=3.5 Myr (top right), and for MHD clouds at $t_{\rm evol}$=2 Myr (bottom left) and $t_{\rm evol}$=3.5 Myr (bottom right). A negative value means that the structure has overall a radial velocity towards its centre of mass (i.e. inflowing). The color-bar represents the boundness of the structures, with reddish structures representing $\alpha_{\mathrm{vir}}^{\mathrm{vol}}<2$. The bound structures ($\alpha_{\mathrm{vir}}^{\mathrm{vol}}<1$), corresponding to points below the horizontal black bar in the color bar) are marked with black outlines. Large scale marginally bound structures show signs of inflowing velocity, } 
    \label{fig:inflow}
\end{figure*}

The GHC scenario suggests that the high line-widths seen in MCs are a result of infalling gas. On the other hand, if gas is swept up due to larger scale flows, it could also result in possible signs of inflowing gas. To investigate this, we compute the radial velocity as
\begin{equation}
    v_r = \frac{1}{M}\int_V \rho (\mathbf{v} - \mathbf{v}_0)\cdot\frac{(\mathbf{r}-\mathbf{r}_0)}{|\mathbf{r}-\mathbf{r}_0|}\mathrm{d}^3r,
\end{equation}
where $\mathbf{r}_0$ is the position of the centre of mass of a given structure. We show the ratio of the radial velocity of the structures over their 1-dimensional velocity dispersion against $R$ in Fig. \ref{fig:inflow}. The color bar here represents $\alpha_{\mathrm{vir}}^{\mathrm{vol}}$, with bound structures ($\alpha_{\mathrm{vir}}^{\mathrm{vol}}<1$) marked with an additional black outline. The top row corresponds to all sub-structures contained in the two hydrodynamic clouds at $t_{\rm evol}=2$ (left) and 3.5~Myr (right), while the bottom row represents the MHD sub-structures. A negative value of $v_r$ suggests that the gas inside the structure has motion towards the centre of mass while a positive value suggests the reverse. If the high kinetic energy in our dendrogram structures originates from inward motions due to gravitational collapse, we would expect more bound structures to have lower (more negative) $v_r/\sigma_{\rm 1D}$ values. From Fig. \ref{fig:inflow}, we do indeed see that the marginally bound points (reddish) tend to have a negative velocity, especially at larger scales, although the trend is far from clear. \par

In order to investigate which effect is dominating (gravitational infall or large-scale, compressive inflows), we perform a correlation analysis between $\alpha_{\mathrm{vir}}^{\mathrm{vol}}$ and $v_r/\sigma_{\rm 1D}$. This can be seen in Table~\ref{tab:infall_kendall}, where we compute the Kendall $\tau$ parameter at 2 and 3.5 Myr for all (both HD and MHD) large-scale ($R>10$~pc) as well as bound and marginally bound ($\alpha_{\rm vir}^{\rm vol}<2$) sub-structures. In each case, we give the Kendall $\tau$ coefficient, k$_{\tau}$, and the p-value. k$_{\tau}$ quantifies the rank correlation between two quantities: +1 denotes perfect positive correlation and -1 perfect negative correlation \citep[][]{kendall_new_1938}. \par 
\begin{table*}
\centering
\begin{tabular}{ccccccccc}
\hline
     $t_{\rm evol}$& \multicolumn{2}{c}{ all}&&\multicolumn{2}{c}{$R$>10 pc}&&\multicolumn{2}{c}{$\alpha_{\rm vir}^{\rm vol}<2$}\\
      \cline{2-3} \cline{5-6} \cline{8-9}
     [Myr]& k$_{\tau}$&p-value&&k$_{\tau}$&p-value&&k$_{\tau}$&p-value\\
     \hline
     2&0.19&3$\times 10^{-9}$&&0.11&0.46&&-0.14&0.009\\
     3.5&0.21&1.1$\times10^{-24}$&&-0.43&1.1$\times10^{-4}$&&-0.07&0.03\\
\hline     
\end{tabular}
    \caption{The Kendall $\tau$ correlation values between $v_r/\sigma_{\rm 1D}$ and $\alpha_{\mathrm{vir}}^{\mathrm{vol}}$ for all, large scale, and bound structures at different times. In each case, the $\tau$ statistic, $k_{\tau}$, and the corresponding p-value is shown. Overall, the structures show a weak correlation between $v_r/\sigma_{\rm 1D}$ and $\alpha_{\mathrm{vir}}^{\mathrm{vol}}$, i.e. a larger $\alpha_{\mathrm{vir}}^{\mathrm{vol}}$ leads to a larger outwards velocity. However, the large scale structures experiencing inflow show an anti-correlation at $t_{\rm evol}=3.5$~Myr, i.e. less bound structures on the large scales have a higher inflow velocity, suggesting compression.}
    \label{tab:infall_kendall}
\end{table*}
Firstly, we note that for \textit{all} structures, there is a statistically significant, mild positive correlation (0.19 at 2 Myr and 0.21 at 3.5 Myr) between $v_r/\sigma_{\rm 1D}$ and $\alpha_{\rm vir}^{\rm vol}$. Since most of the sub-structures are unbound, this reflects rather the fact that structures with higher $\alpha_{\rm vir}^{\rm vol}$ tend to have higher dispersive motions. \par
For large-scale structures, at 2 Myr, we do not find a statistically significant correlation (p-value of 0.46). However, at 3.5 Myr, there is a relatively strong anti-correlation (k$_{\tau}=-0.43$). This implies that a more negative $v_r/\sigma_{\rm 1D}$ is associated with a higher $\alpha_{\rm vir}^{\rm vol}$, i.e. at large scales the more unbound structures tend to feel a stronger inwards motion. This is contrary to our expectations of a gravitational infall, and rather suggest an inflow, where the stronger compression is associated with a higher kinetic energy, and consequently, a higher virial ratio. \par
Over all structures with $\alpha_{\rm vir}^{\rm vol}<2$, we find an even milder or almost no correlation at both earlier and later times (k$_{\tau}$=-0.14 at 2 Myr and -0.07 at 3.5 Myr). The mild anti-correlation here is likely a result of the anti-correlation seen for the larger scale structures, which are mostly marginally bound.

\subsection{Gravo-turbulence vs global hierarchical collapse}
The gradual formation of dense, gravity dominated branches in Larson's $\sigma_{\rm 1D}$-$R$ relation (Fig. \ref{fig:sigma_size}, see also Appendix \ref{sec:A0}), the emergence of gravitationally bound structures over time (Figs. \ref{fig:HD_energy} and \ref{fig:MHD_energy}, bottom panel), the decrease in $\alpha_{\rm vir}^{\rm vol}$ for large-scale unbound and marginally bound structures (Table~\ref{tab:alpha_large_scale}), and the significant negative correlation between $v_r/\sigma_{\rm 1D}$ and $\alpha_{\rm vir}^{\rm vol}$ at large scales (Table~\ref{tab:infall_kendall}) suggest that the dense molecular gas forms from the rarefied medium being swept up by large scale flows, likely originating from the supernovae being driven in the original SILCC simulations. We therefore conclude that our findings support the GT over GHC scenario of structure formation in our simulated clouds. Our results are in line with the new review by \citet{hacar_initial_2022} on filamentary structures and their fragmentation, which provides independent evidence against GHC. In our companion paper (Ganguly et al., in prep.), we suggest that the MCs in our simulations rather form as sheet-like structures tracing the shells of expanding or interacting supernova bubbles, in agreement with the bubble-driven filament formation scenario of MCs \citep[][]{koyama_origin_2002, inoue_two-fluid_2009, inutsuka_formation_2015}.\par
In similar simulations of forming MCs in a supernova-driven, stratified, turbulent medium, \citet{ibanez-mejia_feeding_2017} have found that their clouds remain overall gravitationally bound, and compressions due to supernovae are insufficient to drive the turbulence in the dense molecular gas. This is in contrast to our results. While it is difficult to compare the simulations directly, we hypothesise that the difference between the results could lie in the technique of forming dense structures. \citet{ibanez-mejia_feeding_2017} have no self-gravity in the beginning and model gravitational interactions after the inception of the clouds. In contrast, our clouds include self-gravity from the very beginning and therefore form more self-consistently. This perhaps suggests that modelling gravitational interactions, including tidal effects, in the rare gas is important for consistently modelling the velocity structures and dynamics of the  denser sub-structures that form later on. Once the dense, molecular gas has been formed, additional external nearby supernovae (at distances less than 50~pc) only drive turbulence for a short time \citep[][]{seifried_is_2018}. \par
The role of supernovae in creating a turbulent cascade has been extensively discussed as a possible origin of the Larson-like scaling relations in simulated clouds \citep[e.g.][]{kritsuk_supersonic_2013, padoan_supernova_2016}. The role of the combined effect of supernova-driven turbulence and self-gravity has also been suggested with relation to the observed relations \citep[see e.g.][]{izquierdo_cloud_2021}. Our findings are in agreement with these results.

\section{Internal vs external gravity}\label{sec:gravity}
\subsection{Comparison of acceleration terms}

In the analysis done so far, we have concentrated only on the self-gravity of individual structures. Due to the chaotic nature of the medium, the clumpy surrounding medium contributes to an asymmetric gravitational pull on any given structure, and thereby influences its dynamics. The gravity of the surrounding medium causes tidal effects within each structure, which can result in both, shear motions to tear a structure apart, as well as tidal compression to enhance its collapse along certain directions. Note that tidal compression due to the external medium can never be compressive from all directions, and rather deforms the mass distribution. \par
\begin{figure}
    \centering
    \includegraphics[width=0.49\textwidth]{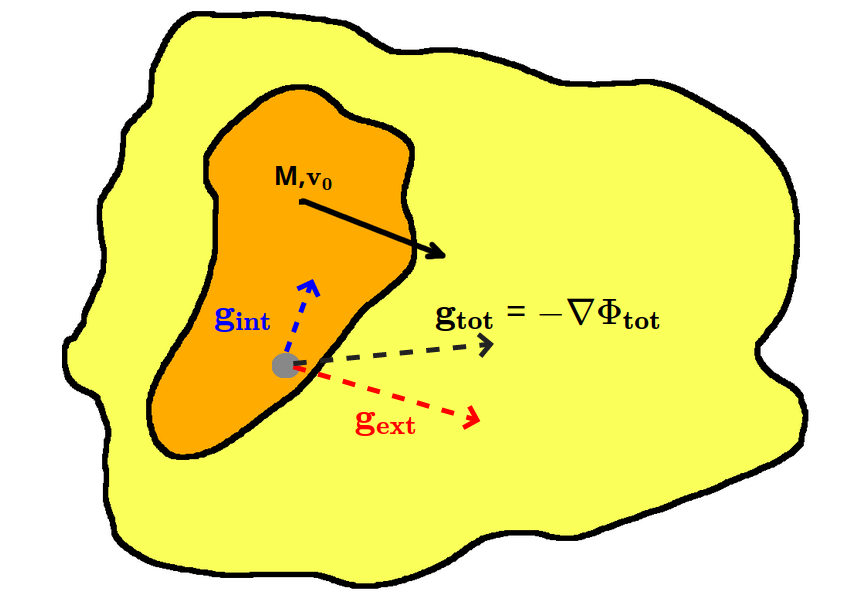}
    \caption{Sketch representing the internal and external gravitational acceleration. A given structure mass M and centre of mass velocity $\mathbf{v}_0$ is embedded in a more diffuse surrounding medium. For each point inside the structure, we compute the net gravitational acceleration $\mathbf{g_{tot}}$ as the negative gradient of the global potential $\Phi_{\rm tot}$, and the self gravitational acceleration $\mathbf{g_{int}}$ based on the mass distribution of the structure itself using a KD tree. The effect of solely the surrounding medium is then the vectorial subtraction of these two terms according to Eq. \ref{eq:gext}.}
    \label{fig:cartoon}
\end{figure}

We highlight the relation between the acceleration due to self-gravity and the surrounding medium in Fig. \ref{fig:cartoon}. We consider a structure embedded in a more diffuse medium. For each point inside this structure, the total gravitational acceleration, $\mathbf{g}_{\rm tot}$, can be computed as
\begin{equation}
    \mathbf{g}_{\rm tot}=-\nabla \Phi_{\rm tot},
\end{equation}
where $\Phi_{\rm tot}$ is the global gravitational potential due to all of the gas in the simulation and is obtained directly from the simulation data. We estimate the acceleration due to self-gravity, $\mathbf{g}_{\rm int}$, by a KD tree algorithm using only the mass contained inside the structure. We can then obtain the acceleration solely due to the surrounding medium by the vector subtraction of these two quantities.
\begin{equation}\label{eq:gext}
    \mathbf{g}_{\rm ext}=\mathbf{g}_{\rm tot}-\mathbf{g}_{\rm int}
\end{equation}
Note that $\mathbf{g}_{\rm tot}$ (and as a consequence $\mathbf{g}_{\rm ext}$) here is caused by the gas distribution, and does not include the galactic potential. The scale height of the galactic potential due to old stars in the simulations is 100~pc, much larger than our largest structures. We therefore neglect its effect. \par 
If self-gravity is the only important term, then the $\mathbf{g}_{\rm int}$ and $\mathbf{g}_{\rm tot}$ terms should be comparable, and the angle between them should peak at a value close to 0. The large-scale structures should also peak near zero, by virtue of the fact that there simply is little gas outside the structures to exert significant tidal forces. \par 

\begin{figure}
    \centering
    \includegraphics[width=0.49\textwidth]{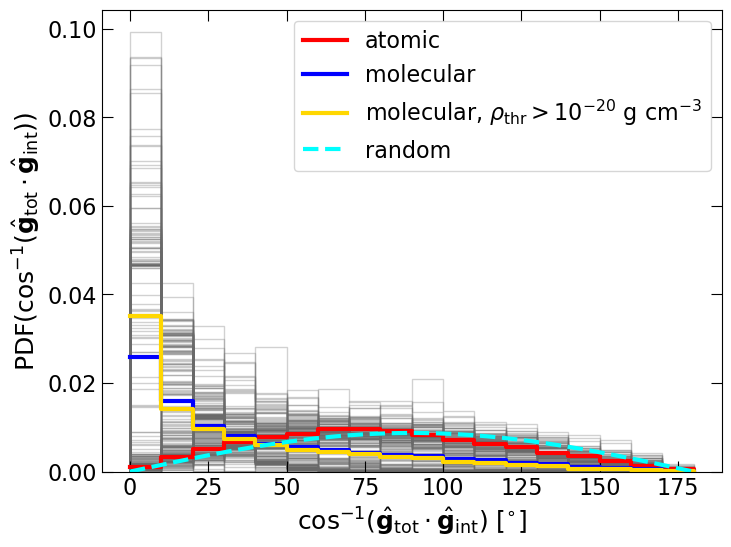}
    \caption{Histogram of angles between the self gravitational acceleration $\mathbf{g}_{\rm int}$ and the total gravitational acceleration $\mathbf{g}_{\rm tot}$ due to gas both inside and outside the structure, for all different dendrogram structures of the cloud HD 1 at $t_{\rm evol}=$3.5 Myr. The histogram for each individual structure is plotted with a grey line. The cyan dashed line shows how the distribution should be if these two vectors were randomly aligned (i.e. if self-gravity plays no role). The average for the atomic, molecular, and dense molecular structures is shown in red, blue, and yellow, respectively. We see that a preferential alignment between the two vectors is seen prominently for molecular structures.}
    \label{fig:gint_got_angle}
\end{figure}

We show the distribution of angles between these two terms, denoted by cos$^{-1}(\hat{\mathbf{g}}_{\rm tot}\cdot\hat{\mathbf{g}}_{\rm int})$, as PDF computed over all cells for each dendrogram structure of one example cloud (MC1-HD at 3.5 Myr) in Fig. \ref{fig:gint_got_angle}. Here, $\hat{\mathbf{g}}_{\rm tot}$ and $\hat{\mathbf{g}}_{\rm int}$ are the unit vectors of $\mathbf{g}_{\rm tot}$ and $\mathbf{g}_{\rm int}$, respectively, i.e.
\begin{gather}
    \hat{\mathbf{g}}_{\rm tot}=\frac{\mathbf{g}_{\rm tot}}{|\mathbf{g}_{\rm tot}|},\
    \hat{\mathbf{g}}_{\rm int}=\frac{\mathbf{g}_{\rm int}}{|\mathbf{g}_{\rm int}|}.
\end{gather}
The histogram of each individual dendrogram structure is shown by a grey line. If the angle between these two vectors is completely random, as could be the case if self-gravity is not at all important, then the histogram of angles should follow a distribution $\mathrm{P}(\theta) \propto \mathrm{sin}(\theta)$. This is denoted by the dashed cyan line. The average behaviour for atomic, molecular, and dense molecular ($\rho_{\rm thr}>10^{-20}$~g~cm$^{-3}$) sub-structures for this particular cloud is shown in red, blue and yellow, respectively. \par 

From the individual grey outlines, we see that for many structures, the angle between the two vectors is not random, and there is some preferential alignment. A few of the structures do indeed peak near 0, while there are also a significant number of structures where the angle distribution is relatively random. A clear distinction is seen when calculating the average behaviour for the molecular structures and the atomic structures, separately. The atomic structures show a behaviour which is in good agreement with the random line, suggesting that they are only marginally affected by their self-gravity. In contrast, the average behaviour of the molecular structures clearly shows that self-gravity plays a more pivotal role in their evolution. \par
\begin{figure*}
    \centering
    \includegraphics[width=0.49\textwidth]{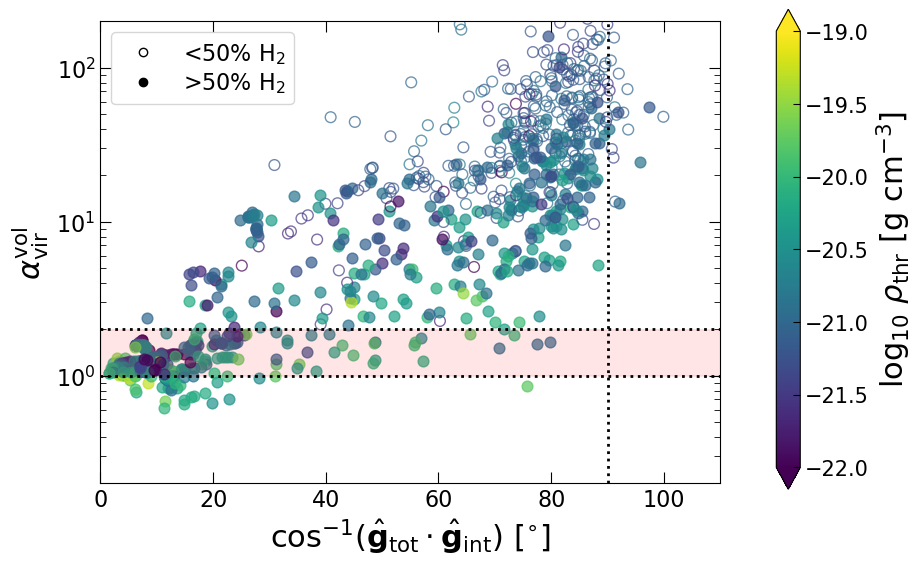}
    \includegraphics[width=0.49\textwidth]{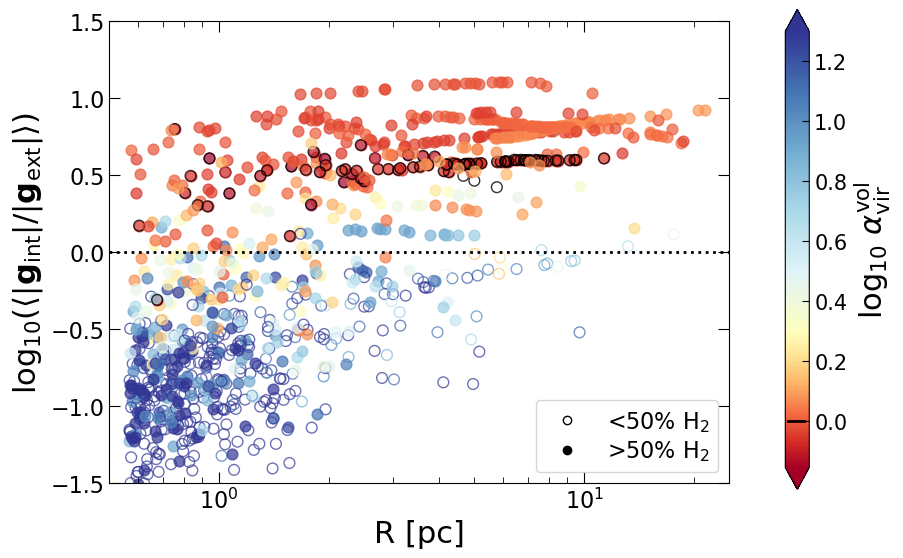}
    \caption{Left: The virial parameter $\alpha_{\mathrm{vir}}^{\mathrm{vol}}$ of the structures, plotted against the median angle distribution between the self gravitational acceleration $\mathbf{g}_{\rm int}$ and the total gravitational acceleration $\mathbf{g}_{\rm tot}$, for all clouds, both HD and MHD at $t_{\rm evol}=$3.5 Myr. The red shaded region shows the region between marginally bound and virialized, while the vertical dotted line shows the expected value for a completely random alignment between $\mathbf{g}_{\rm int}$ and $\mathbf{g}_{\rm tot}$. More bound structures have a preferential alignment between $\mathbf{g}_{\rm int}$ and $\mathbf{g}_{\rm tot}$. Right: Average ratio of $\mathbf{g}_{\rm int}$ to the external gravitational acceleration $\mathbf{g}_{\rm ext}$, plotted for each structure against its size, for all clouds at $t_{\rm evol}=$3.5 Myr. The colorbar shows the virial ratio $\alpha_{\mathrm{vir}}^{\mathrm{vol}}$, and gravitationally bound structures are marked with a black circle. For the unbound, and mostly but not exclusively atomic structures, the external gravity seems to be more important compared to the self-gravity of the structures.} 
    \label{fig:angle_median}
\end{figure*}

We find consistent results when looking at the median angle distribution against $\alpha_{\mathrm{vir}}^{\mathrm{vol}}$, for all the different cloud structures in Fig. \ref{fig:angle_median} (left panel). The median angle for a given sub-structure here is computed by obtaining the median for cos$^{-1}(\hat{\mathbf{g}}_{\rm tot}\cdot\hat{\mathbf{g}}_{\rm int})$ over all cells in the structure. The horizontal red band represents the marginally bound region ($1\leq \alpha_{\mathrm{vir}}^{\mathrm{vol}}<2$), while the vertical dotted line represents the expected median for a completely random alignment, i.e. 90 degrees. We see a clear correlation between the median angle and the virial ratio, with smaller median angles corresponding to smaller virial ratios. Interestingly, there seem to be some structures, which are marginally bound but still have a relatively high median angle. These could possibly represent structures that are at least partially compressed due to tidal forces.\par

We find a similar picture if we compare the density weighted mean of the ratio of magnitudes of $\mathbf{g}_{\rm int}$ and $\mathbf{g}_{\rm ext}$ for each structure. This is shown in the right panel of Fig. \ref{fig:angle_median}, where we plot the log of the average of $\frac{|\mathbf{g}_{\mathrm{int}}|}{|\mathbf{g}_{\mathrm{ext}}|}$ against $R$. We compute this by:
\begin{equation}
    \left\langle\frac{|\mathbf{g}_{\mathrm{int}}|}{|\mathbf{g}_{\mathrm{ext}}|} \right\rangle=  \frac{1}{M}\int_V \frac{|\mathbf{g}_{\mathrm{int}}|}{|\mathbf{g}_{\mathrm{ext}}|}\rho \mathrm{d}^3r
\end{equation}
The colorbar here represents $\alpha_{\mathrm{vir}}^{\mathrm{vol}}$, with bound structures ($\alpha_{\mathrm{vir}}^{\mathrm{vol}}<1$) marked with an additional black outline. We find that large scale (>10 pc) structures experience only a small effect from external gravity, owing to the fact that they represent almost the entire cloud and there is little mass outside to impart significant tidal influence. At smaller scales, we do continue to have structures where self-gravity is far more important compared to the gravity of the surrounding medium, but we also find a large number of (predominantly atomic, but also molecular) structures where the external gravity can play an important role. There is also a clear trend that more bound structures tend to have stronger self-gravity, and therefore experience less influence due to the external medium. It is important to note that this analysis compares only gravity terms, and therefore does not tell us about the overall importance of gravity compared to other forces (see Section \ref{sec:energetics}).\par
\begin{figure*}
    \centering
    \includegraphics[width=0.49\textwidth]{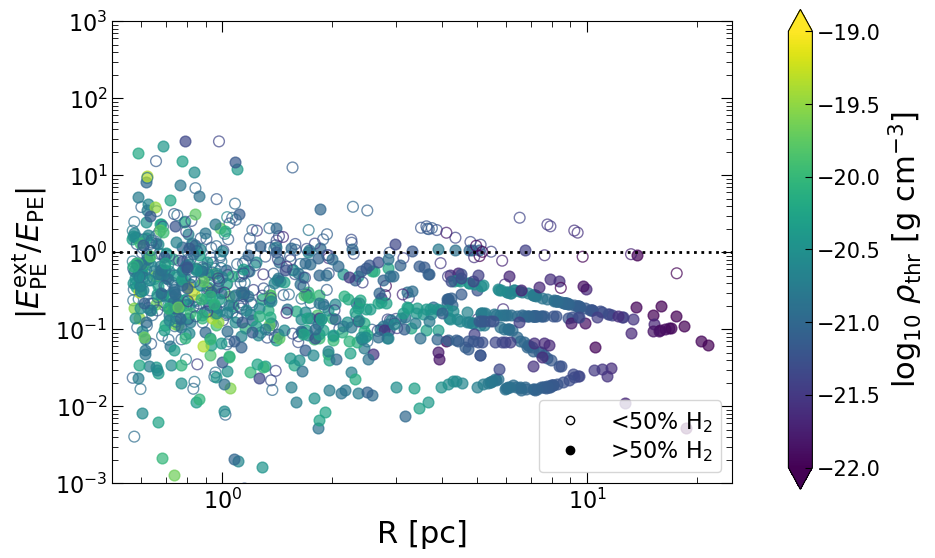}
    \includegraphics[width=0.49\textwidth]{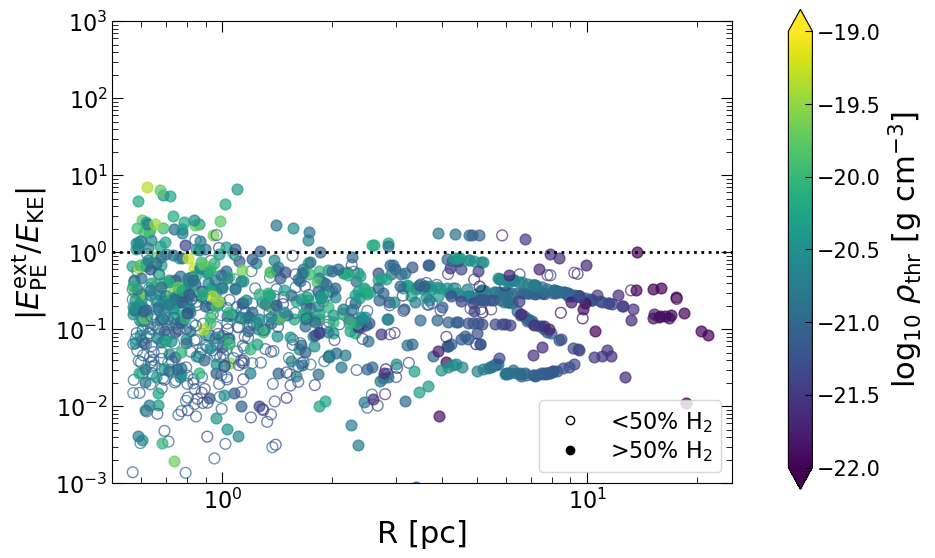}
    \caption{Ratio of the magnitudes of external gravitational energy $E_{\mathrm{PE}}^{\mathrm{ext}}$ to the self gravitation potential energy (left) and to the kinetic energy (right) for all clouds, both HD and MHD, at $t_{\rm evol}=$3.5 Myr. The potential energy due to the external medium is generally less for almost all structures compared to both potential and kinetic energy. This, however, masks compression and extension happening to the same structure due to the external medium, which would cancel out in the energy term but still deform the structure.}
    \label{fig:eratio_gext}
\end{figure*}

Finally, we attempt to directly compute the energy associated with the external gravitational field $\mathbf{g}_{\mathrm{ext}}$. We estimate this from
\begin{equation}
    E_{\mathrm{PE}}^{\mathrm{ext}} = \int_V \rho (\mathbf{r}-\mathbf{r}_0)\cdot\mathbf{g}_{\mathrm{ext}} \;\, \mathrm{d}^3r.
\end{equation}
The comparison of $E_{\mathrm{PE}}^{\mathrm{ext}}$ with the self-gravitating potential energy and kinetic energy at $t_{\rm evol}=3.5$~Myr is seen in Fig. \ref{fig:eratio_gext} for all sub-structures, both HD and MHD. The x-axis represents the size $R$ of structure, while the colorbar plots the density threshold. The horizontal dotted line represents a value of 1 in the ratio for both panels. We see that $E_{\mathrm{PE}}^{\mathrm{ext}}$ is smaller than the kinetic (right panel) or the self-gravitating potential energy (left panel) for almost all structures. This, however, hides the potentially dynamic effect tidal forces can have. If a structure is being partially compressed on one side and torn apart on the opposite side, in terms of energy these two effects would tend to cancel each other but the structure is actually being deformed. This can be captured by performing a more nuanced tidal analysis, which we do in the following. 

\subsection{Tidal analysis}\label{sec:tidal_tensor}
The gravity of the external medium is important for a significant number of structures in our clouds. It is, however, unclear what exact effect the external gravity has. The structures in the simulations live in a chaotic environment, and therefore the net effect of the tides can cause shear motions, disrupting their possible collapse or hindering compression. Conversely, tidal effects may cause at least partial compression. To quantify the effect of the gravity of the medium surrounding the structure, we perform an analysis based on the tidal tensor \citep[see e.g.][]{renaud_fully_2009, renaud_evolution_2011}. \par 
For a gravitational potential field $\Phi$, the tidal tensor $\mathbf{T}$ is defined such that
\begin{equation}
    \mathrm{T}_{ij} = -\partial_i \partial_j \Phi.
\end{equation}
The eigenvalues of the tidal tensor encode the information related to deformation (compression/extension) occurring due to gravity at a certain point due to the local gravitational field. Because the tidal tensor is symmetric and real-valued, it can be written in orthogonal form. The diagonalized tidal tensor is given as, 
\begin{gather}
    \mathbf{T}
    =
    \begin{bmatrix}
    \lambda_1 &0&0\\
    0&\lambda_2&0\\
    0&0&\lambda_3
    \end{bmatrix}
\end{gather}
where $\lambda_i$ are its eigenvalues. The trace of the tidal tensor $\mathrm{Tr(\mathbf{T})}=\sum_{i=1}^{3} \lambda_i$ contains the local density information in the Poisson equation:
\begin{equation}
    \mathrm{Tr(\mathbf{T})} = -\nabla^2 \Phi = -4\pi G \rho.
\end{equation}
This implies that the trace is 0 if the point at which the tidal tensor is evaluated is outside the mass distribution. \par
For example, for a point of mass $M$, the tidal tensor at distance $x$ is given as:
\begin{gather}
    \mathbf{T}
    =
    \begin{bmatrix}
    \frac{2GM}{x^3} &0&0\\
    0&-\frac{GM}{x^3}&0\\
    0&0&-\frac{GM}{x^3}
    \end{bmatrix}
\end{gather}
The sign of the eigenvalues indicates whether the given mode is compressive or extensive, while the magnitude represents the strength of the respective compressive/extensive mode \citep[see e.g.][]{renaud_fully_2009}. A positive eigenvalue $\lambda_i > 0$ implies that a clump of gas will expand along that particular eigen-direction due to the local gravitational field, while a negative eigenvalue of $\lambda_i < 0$ implies the opposite. This does not of course automatically imply that the structure will successfully collapse or disperse along the given eigen-direction. The present analysis is done only on the gravitational field, and reflects the nature of gravitational deformation, in absence of all other resisting forces such as thermal or magnetic pressure.\par
For the purpose of our analysis, we are interested if an entire structure will compress/extend due to the gravitational field. The average tidal tensor for an entire structure can be computed as the volume average of the tidal tensor at each point:
\begin{equation}
    \langle \mathrm{T} \rangle_{ij} = \frac{1}{V} \int_V \partial_i g_j \mathrm{d}^3r.
\end{equation}
Similar to splitting the gravitational acceleration vector in the previous section, we can split the tidal tensor also into three parts: the tidal tensor due to only the matter inside the structure, $\mathbf{T}_{\mathrm{int}}$, due to only matter distribution external to the structure, $\mathbf{T}_{\mathrm{ext}}$, and their sum due to the entire matter distribution:
\begin{equation}
\langle\mathbf{T}\rangle_{\mathrm{tot}}=\langle\mathbf{T}\rangle_{\mathrm{int}}+\langle\mathbf{T}\rangle_{\mathrm{ext}}.   
\end{equation}
Here, $\mathrm{\langle\mathbf{T}\rangle_{tot}}$ represents the net deformation due to both, the structure itself and its surroundings, while $\mathrm{\langle\mathbf{T}\rangle_{ext}}$ represents the deformation introduced solely due to the external medium. The trace of the three different tidal tensors are as follows:
\begin{gather}\label{eq:trace1}
    \mathrm{Tr(\langle\mathbf{T}\rangle_{tot})} = -4\pi G \rho_{\mathrm{avg}}\\
    \mathrm{Tr(\langle\mathbf{T}\rangle_{int})} = -4\pi G \rho_{\mathrm{avg}}\\
    \mathrm{Tr(\langle\mathbf{T}\rangle_{ext})} = 0,\label{eq:trace3}
\end{gather}
where $\rho_{\mathrm{avg}}$ is the average density computed in Eq. \ref{eq:avg_den}. We show in the Appendix \ref{sec:A1} that the traces are retrieved. An important consequence of Eqs. \ref{eq:trace1}-\ref{eq:trace3} is that $\mathrm{\langle\mathbf{T}\rangle_{ext}}$ must contain both compressive and extensive (disruptive) modes, corresponding to negative and positive $\lambda_i$ respectively. In contrast, $\mathrm{\langle\mathbf{T}\rangle_{int}}$ and $\mathrm{\langle\mathbf{T}\rangle_{tot}}$ must contain at least one compressive mode, but might also be fully compressive. \par
\begin{figure}
    \centering
    \includegraphics[trim=8 5 3 2, clip,width=0.49\textwidth]{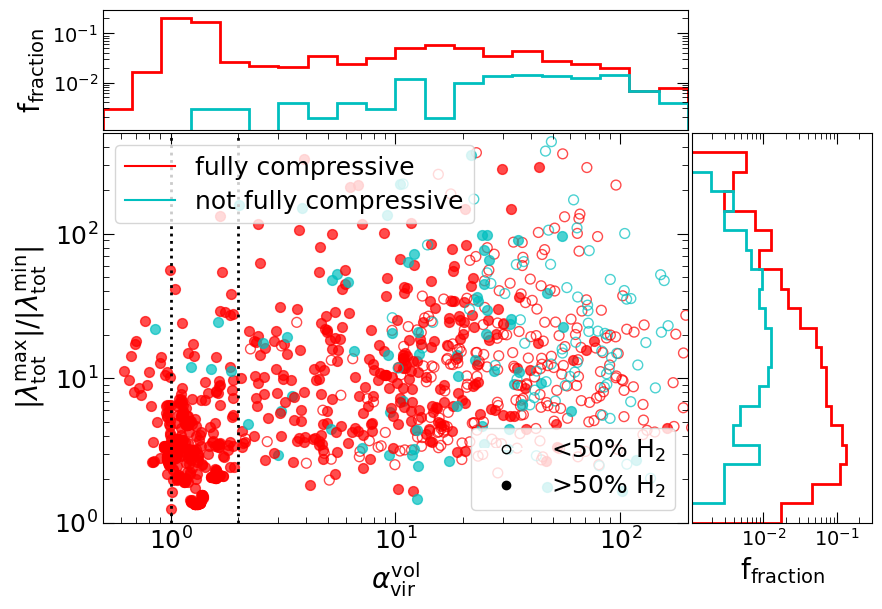}
    \caption{The ratio of the largest to the smallest eigenvalue of the tidal tensor $\langle\mathbf{T}\rangle_{\mathrm{tot}}$ plotted against $\alpha_{\mathrm{vir}}^{\mathrm{vol}}$ of the structures, for all clouds at $t_{\rm evol}=$3.5 Myr. The two vertical lines denote the marginally bound region ($1\leq\alpha_{\mathrm{vir}}^{\mathrm{vol}}<2$). The red symbols represent fully compressive structures (all $\lambda_i<0$), while the cyan ones are structures which have at least extensive mode. The side-panes show the 1D distribution of the fraction of structures based on the distribution along that axis, and are normalized by the total number of structures. The $\lambda_{\mathrm{tot}}^{\mathrm{max}}/\lambda_{\mathrm{tot}}^{\mathrm{min}}$ ratio represents the degree of anisotropy in the gravitational field. For most structures, the overall deformation due to gravity is still fully compressive, suggesting that disruptive tidal effects are not dominant. Structures with at least one extensive mode seem to experience a more anisotropic (higher $|\lambda_{\mathrm{tot}}^{\mathrm{max}}|/|\lambda_{\mathrm{tot}}^{\mathrm{min}}|$) gravitational field. They are also more likely to be more unbound, while marginally or gravitationally bound structures are almost always fully compressive, with a few exceptions.}
    \label{fig:lambda_gtot}
\end{figure}

In Fig. \ref{fig:lambda_gtot}, we plot the ratio of the maximum to the minimum absolute eigenvalue of the total tidal tensor $|\lambda_{\mathrm{tot}}^{\mathrm{max}}|/|\lambda_{\mathrm{tot}}^{\mathrm{min}}|$ against the virial ratio of the structures. These are computed as
\begin{gather}\label{eq:lambdamax}
    |\lambda_{\mathrm{tot}}^{\mathrm{max}}| = \mathrm{max}(\{|\lambda_{\mathrm{i,tot}}|\})\\
    |\lambda_{\mathrm{tot}}^{\mathrm{min}}| = \mathrm{min}(\{|\lambda_{\mathrm{i,tot}}|\}), i=1,2,3.
\end{gather}
Since the eigenvalues can become negative, we consider their absolute values in order to disentangle the relative importance of tidal compression ($\lambda<0$) vs extension ($\lambda>0$). Fully compressive deformations (all three $\lambda_i<0$) are shown in red, while those with at least one extensive mode are shown in cyan. The two vertical dashed lines represent $\alpha_{\mathrm{vir}}^{\mathrm{vol}}=1$ and $\alpha_{\mathrm{vir}}^{\mathrm{vol}}=2$. The top and right side-panels show the fraction ${\rm f_{fraction}}$ of structures against $\alpha_{\mathrm{vir}}^{\mathrm{vol}}$ and $|\lambda_{\mathrm{tot}}^{\mathrm{max}}|/|\lambda_{\mathrm{tot}}^{\mathrm{min}}|$, respectively, for both fully compressive and partially extensive structures.  \par

We first note that most structures have three compressive modes, suggesting that the inclusion of the surrounding clumpy medium, while modifying the nature of gravitational interaction, still results in compressive deformation due to the self-gravity. Note that this analysis is done on the gravity alone, and does not therefore imply that the structure is actually collapsing, only that this is the net effect the overall gravitational field will achieve. \par
We further find that the compression due to gravity is highly anisotropic, and on average appears to be more anisotropic (higher $|\lambda_{\mathrm{tot}}^{\mathrm{max}}/\lambda_{\mathrm{tot}}^{\mathrm{min}}|$) for structures with at least one extensive mode (Fig. \ref{fig:lambda_gtot}, right panel histogram). A higher anisotropy indicates that gravity is attempting to flatten or elongate these structures. Partially extensive structures also seem to experience such an elongating effect more strongly compared to fully compressive structures. \par 
From Fig. \ref{fig:lambda_gtot}, top panel histogram, we also see that structures which have at least one extensive mode, also tend to be more unbound. This raises the intriguing possibility that perhaps their high $\alpha_{\mathrm{vir}}^{\mathrm{vol}}$ values are even generated by the highly anisotropic deformation introduced due to tides, by converting tidal energy into kinetic energy. \par  
To investigate this possible scenario, and to quantify the relevance of the tidal force compared to other terms (such as self-gravity and turbulence), we perform a further timescale analysis based on the tidal tensor. For this purpose, we are interested in quantifying the relevant timescales and energies introduced by the external medium, and therefore use $\mathrm{\langle\mathbf{T}\rangle_{ext}}$ for our analysis. \par
The eigenvalues of the tidal tensor have units of [time]$^{-2}$, and can therefore directly be converted into a timescale. Assuming that the largest eigenvalue dominates the deformation of a given structure, we can define a tidal timescale which represents the typical deformation timescale of the given structure solely due to the external gravitational field as follows: 
\begin{equation}
    t_{\mathrm{tidal}}^{\mathrm{ext}} = (|\lambda_{\mathrm{ext}}^{\mathrm{max}}|)^{-1/2},
\end{equation}
where $|\lambda_{\mathrm{ext}}^{\mathrm{max}}|$ is the maximum of the absolute eigenvalues $|\lambda_{i,\mathrm{ext}}|$, similar to Eq. \ref{eq:lambdamax}.\par
To compare this deformation timescale with the typical timescale of the gravity and turbulence, we use the free-fall time, $t_{\mathrm{ff}}$, and the crossing time, $t_{\mathrm{crossing}}$, respectively. The free fall time represents the timescale over which a uniform density spherical structure would collapse in the absence of any pressure forces, solely due to its own self-gravity:
\begin{equation}
    t_{\mathrm{ff}} = \sqrt{\frac{3\pi}{32G\rho_{\mathrm{avg}}}}.
\end{equation}
If the ratio $t_{\mathrm{tidal}}^{\mathrm{ext}}/t_{\mathrm{ff}} >> 1$, this implies that the self-gravity of the structure dominates and acts on a much shorter timescale compared to any deformation due to external tidal forces. In contrast, if $t_{\mathrm{tidal}}^{\mathrm{ext}}/t_{\mathrm{ff}} << 1$, then tidal deformation is important in terms of modifying any possible collapse of a given structure. \par
Similarly, the crossing timescale is computed as
\begin{equation}
    t_{\mathrm{crossing}} = \frac{c}{\sigma_{\rm 1D}},
\end{equation}
where $c$ is the shortest axis of the fitted ellipsoid (Eq.~\ref{eq:fit_ellipsoid}). The crossing timescale represents the timescale over which supersonic turbulence is established throughout the medium. In the absence of driving forces, supersonic turbulence is also expected to decay over $t_{\rm crossing}$ \citep[][]{mac_low_kinetic_1998, stone_dissipation_1998}. If $t_{\mathrm{tidal}}^{\mathrm{ext}}/t_{\mathrm{crossing}} >> 1$, the turbulence dissipates on a much shorter timescale compared to the tidal deformation timescale. On the other hand, if $t_{\mathrm{tidal}}^{\mathrm{ext}}/t_{\mathrm{crossing}} << 1$, the crossing timescale is much longer compared to the tidal timescale, and gravitational deformation occurs on a timescale short enough to possibly generate the high amount of kinetic energy seen in our energy analysis. \par
\begin{figure}
    \centering
    \includegraphics[width=0.49\textwidth]{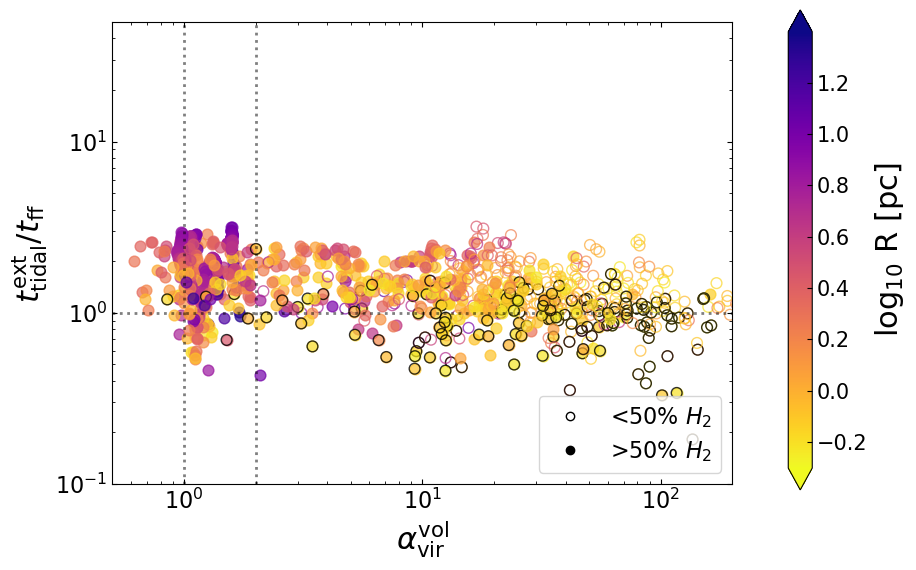}
    \includegraphics[width=0.49\textwidth]{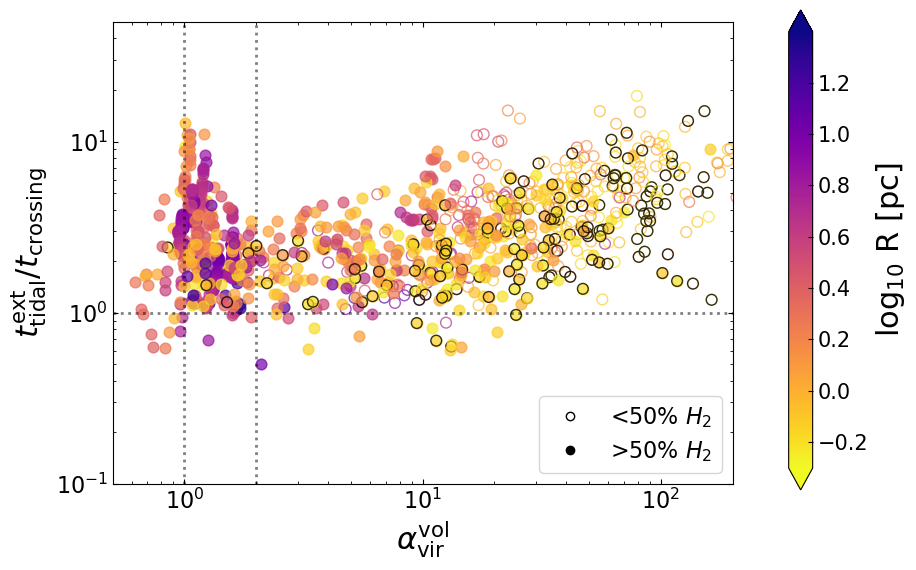}
    \includegraphics[width=0.49\textwidth]{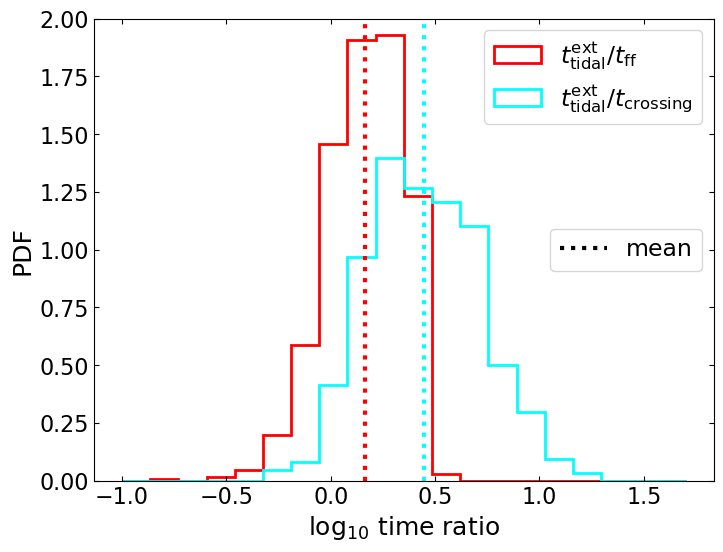}
    \caption{Ratio of the tidal deformation timescale due to $\langle\mathbf{T}\rangle_{\mathrm{ext}}$, to the gravitational free fall timescale (top) and the crossing time (middle), representing the turbulence decay timescale, plotted against the virial parameter $\alpha_{\mathrm{vir}}^{\mathrm{vol}}$ of the substructures, for all clouds at $t_{\rm evol}=$3.5 Myr. The colorbar represents the size of the structures. Structures which have at least one extensive mode (one $\lambda_{i}^{\rm tot}>0$) are marked with a black circle. The vertical dotted lines represent the marginally bound region $1\leq\alpha_{\mathrm{vir}}^{\mathrm{vol}}<2$ for the structures. The top plot shows that self-gravity is becoming more important compared to the tidal deformation timescale on average, for more bound structures. The bottom plot shows that the crossing time is shorter than the tidal deformation timescale for almost all structures, suggesting that tidal deformations due to the surrounding medium cannot be the primary source of turbulent energy. The histogram of the distribution of the two ratios, $t_{\mathrm{tidal}}^{\mathrm{ext}}/t_{\mathrm{crossing}}$ and $t_{\mathrm{tidal}}^{\mathrm{ext}}/t_{\mathrm{ff}}$, with the means plotted in dotted vertical lines (bottom).}
    \label{fig:tratio_gext}
\end{figure}

The behaviour of the two ratios against the virial ratio of the different structures is shown in Fig. \ref{fig:tratio_gext} (top: $t_{\mathrm{tidal}}^{\mathrm{ext}}/t_{\mathrm{ff}}$, middle: $t_{\mathrm{tidal}}^{\mathrm{ext}}/t_{\mathrm{crossing}}$). In both plots, the colorbar represents the size of the structures. The vertical dotted lines show the boundary between virialized and marginally bound. The circles with a black outline indicate structures which have at least one extensive $\lambda_{i}^{\mathrm{tot}}$ (corresponding to the cyan points in Fig. \ref{fig:lambda_gtot}). \par

We see that the ratio $t_{\mathrm{tidal}}^{\mathrm{ext}}/t_{\mathrm{ff}}$ becomes slightly higher with a lower $\alpha_{\mathrm{vir}}^{\mathrm{vol}}$, and therefore self-gravity becomes more and more important as we go to more bound structures. This agrees with the previous picture in Fig.~\ref{fig:angle_median}, where the external gravity was magnitude-wise less important for larger scale, as well as more bound structures. Structures with at least one extensive mode have tidal timescales comparable to or slightly shorter than the free fall time. \par
The ratio $t_{\mathrm{tidal}}^{\mathrm{ext}}/t_{\mathrm{crossing}}$ has quite a different behaviour (Fig. \ref{fig:tratio_gext}, middle panel). For almost all structures, but especially for unbound structures, the crossing timescale is much shorter compared to the tidal timescale. This suggests that turbulence dissipates on a much shorter timescale compared to the gravitational deformation timescale introduced by tides and as such, tides cannot be the principal contributor to the high kinetic energy in the structures. \par
The difference in this average behaviour can be seen in a histogram of these two ratios in Fig. \ref{fig:tratio_gext}, bottom panel. Here, the distribution of $t_{\mathrm{tidal}}^{\mathrm{ext}}/t_{\mathrm{ff}}$ and $t_{\mathrm{tidal}}^{\mathrm{ext}}/t_{\mathrm{crossing}}$ are plotted as a PDF in red and cyan, respectively. The vertical dotted lines represent the mean of each distribution. We see that the distribution of $t_{\mathrm{tidal}}^{\mathrm{ext}}/t_{\mathrm{crossing}}$ is shifted to the right compared to the distribution of $t_{\mathrm{tidal}}^{\mathrm{ext}}/t_{\mathrm{ff}}$, and in both cases the mean and most of the structures lie above a ratio of 1. Overall we find a scenario where $t_{\mathrm{crossing}}<t_{\mathrm{tidal}}^{\mathrm{ext}}$ for most of the structures, while $t_{\rm ff}$ is more comparable to $t_{\mathrm{tidal}}^{\mathrm{ext}}$. This suggests that turbulence acts on a short enough timescale to modify the structures as they evolve.

\section{Conclusions}\label{sec:conclusions}
We perform a detailed energetic analysis of seven different simulated molecular clouds (5 with magnetic fields and 2 without) from the SILCC-Zoom simulations \citep[][]{seifried_silcc-zoom_2017}. In these simulations, we study the evolution of the multi-phase interstellar medium in a supernova-driven, stratified galactic disc environment. We identify structures forming inside each cloud using a dendrogram algorithm, and trace the evolution of their statistical properties over 1.5 Myr during the early stages of cloud evolution before stellar feedback complicates the picture. We include a simple chemical network which allows us to follow the formation of H$_2$ as the cloud assembles and as such, distinguish between mostly atomic (H$_2$ mass fraction < 50\%) and mostly molecular (H$_2$ mass fraction > 50\%) structures. \par
\begin{itemize}
    \item We find that our clouds show a Larson-like power-law behaviour with significant systematic deviations for the densest dendrogram branches in both $\sigma_{\rm 1D}$-size and mass-size relations (Fig.~\ref{fig:sigma_size}). We further see that these deviations evolve over time from a roughly Larson-like law, suggesting that they emerge as gravity becomes more and more important. We observe that all sub-structures can be divided into roughly two kinds depending on their behaviour in the Heyer plot (Fig. \ref{fig:heyer}); denser and molecular structures that roughly follow the Heyer line, and less dense and mostly atomic structures that show no trend with surface density. \par 
    \item In terms of energetics, we find that the dynamics of large scale  ($\geq$ 10 pc) structures, as well as smaller scale denser structures is primarily governed by the interplay between turbulence and gravity (Figs. \ref{fig:HD_energy} and \ref{fig:MHD_energy}). Thermal and magnetic energies are dynamically important only for more diffuse, mostly atomic structures. Our results are based on an analysis of the volume energy terms.
    \item From a virial analysis, we find that on the large scales, the dendrogram structures are unbound or marginally bound at earlier times, but become much closer to virialized over time. Denser, potentially star-forming structures emerge over time, suggesting that they are created by compression in a turbulent, marginally bound environment. \par 
    By performing a correlation analysis between the virial parameter and radial velocity tracing the nfall or inflow for each structure, we find that on the larger scales a more unbound structure is likely to have a stronger inflowing motion. This suggests that we see signs of supernova driven compression, rather than gravitational infall and is in agreement with the bubble-driven filament formation scenario proposed by \citet{inutsuka_formation_2015}. \par
    \item Finally, we attempt to assess the importance of gravitational tidal forces and their role in the dynamics of cloud sub-structures. We compare a given structure's self-gravity and the gravity due to its inhomogeneous surrounding medium by comparing the magnitude of, as well as the angle between, the two respective acceleration terms (Fig.~\ref{fig:angle_median}). We find that the surrounding medium has some influence on a number of smaller scale unbound structures. Energetically, we find that the external gravitational energy is smaller in magnitude compared to both self-gravity and kinetic energy (Fig.~\ref{fig:eratio_gext}). \par
    To quantitatively evaluate the nature of the tidal field, we perform an analysis based on the tidal tensor. We find that the tidal effects are mostly compressive (Fig.~\ref{fig:lambda_gtot}). However, the gravitational field is highly anisotropic and even more so for structures which have extensive modes. Based on a timescale analysis, we find that timescales related to tidal deformation are generally longer compared to the turbulence decay timescale, suggesting that tidal deformations cannot be the primary source of generating kinetic energy in the structures (Fig.~\ref{fig:tratio_gext}). Howeverm the tidal timescale is rather comparable to the freefall time, implying that the structures will deform as they collapse.\par
\end{itemize}\par
Overall, we find a structure formation scenario consistent with the gravo-turbulent scenario of structure formation, with bound structures emerging over time from a largely unbound or marginally bound medium. Magnetic and thermal energy seem to play a subservient role compared to turbulence and self-gravity, with gravitational tides modifying the nature of gravitational compression and leading to formation of more anisotropic, elongated structures. 
\section*{Acknowledgements}
SG, SW and DS would like to acknowledge the support of Bonn-Cologne Graduate School (BCGS), which is funded through the German Excellence Initiative, as well as the DFG for funding through SFB 956 'Conditions and Impact of Star Formation' (subprojects C5 and C6). SDC is supported by the Ministry of Science and Technology (MoST) in Taiwan through grant MoST 108-2112-M-001-004-MY2. This research made use of \texttt{astrodendro} \citep[][]{robitaille_astrodendro_2019}, a Python package to compute dendrograms of Astronomical data (\href{http://www.dendrograms.org/}{http://www.dendrograms.org/}); as well as \texttt{yt}, an open-source, permissively-licensed python package for analyzing and visualizing volumetric data (\href{https://yt-project.org/}{https://yt-project.org/}). The FLASH code used in this work was partly developed by the Flash Center for Computational Science at the University of Chicago.

\section*{Data Availability}
The data underlying this article can be shared for selected scientific purposes after request to the corresponding author.
 



\bibliographystyle{mnras}
\bibliography{2022_01_my_library} 




\appendix
\section{Velocity dispersion-size dependence on the virial ratio}\label{sec:A0}
\begin{figure*}
    \centering
    \includegraphics[width=0.49\textwidth]{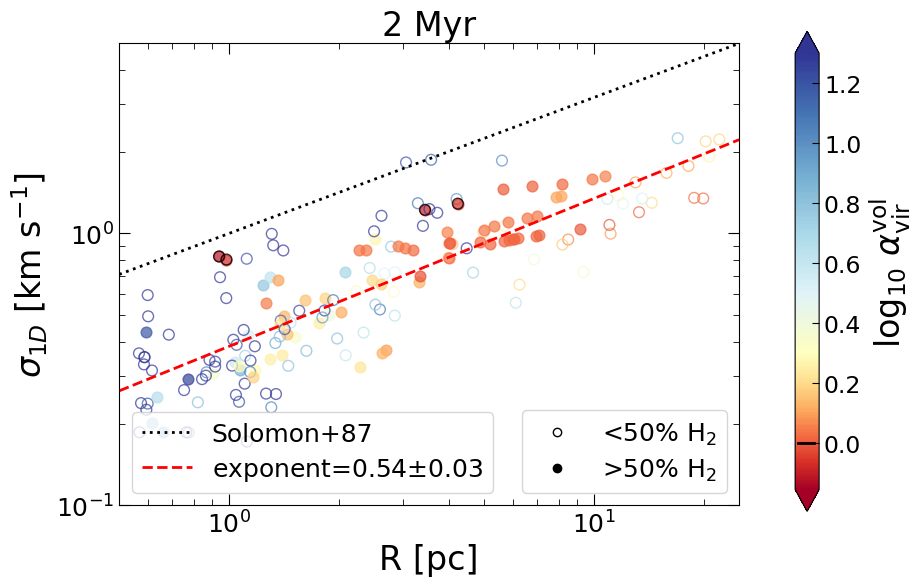}
    \includegraphics[width=0.49\textwidth]{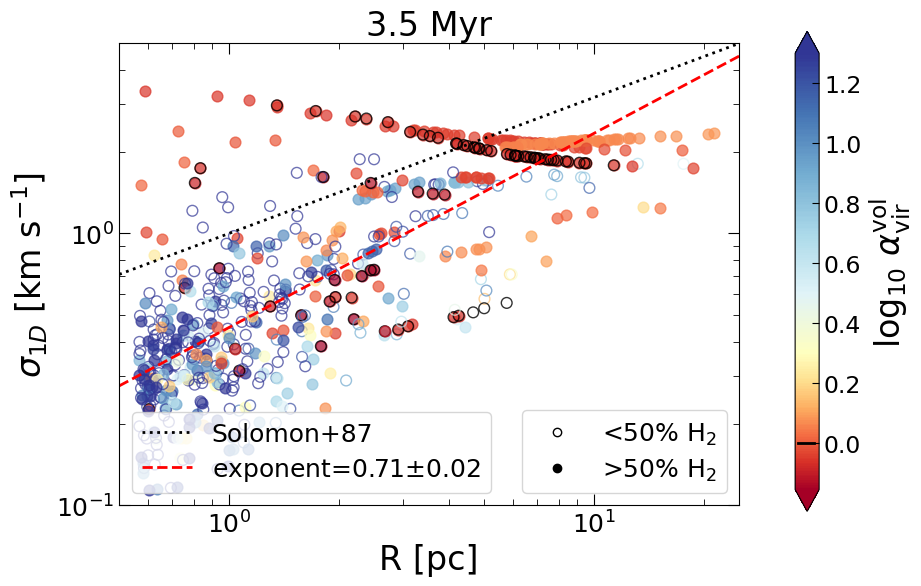}
    \caption{$\sigma_{\rm 1D}$-$R$ relation for all MHD sub-structures, at $t_{\rm evol}=2$ (left) and 3.5 Myr (right). The colour bar represents $\alpha_{\rm vir}^{\rm vol}$, with structures with $\alpha_{\rm vir}^{\rm vol}<1$ being marked with an additional black outline.}
    \label{fig:larson_bound}
\end{figure*}

We highlight the evolution of Larson's velocity dispersion-size relation in relation to $\alpha_{\rm vir}^{\rm vol}$ in Fig. \ref{fig:larson_bound} for all different MHD sub-structures at 2 Myr (Fig. \ref{fig:larson_bound}, left) and 3.5 Myr (Fig. \ref{fig:larson_bound}, right). We see that at 3.5 Myr structures deviating to form flatter branches are all bound or close to bound. However, the deviations are largely absent at earlier times.

\section{Trace of the tidal tensor}\label{sec:A1}
\begin{figure*}
    \centering
    \includegraphics[width=0.49\textwidth]{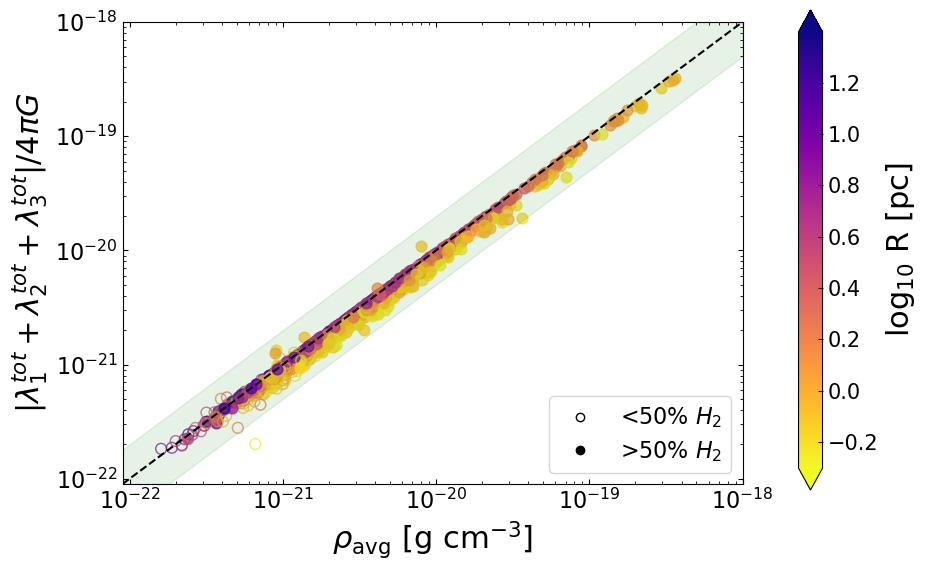}
    \includegraphics[width=0.49\textwidth]{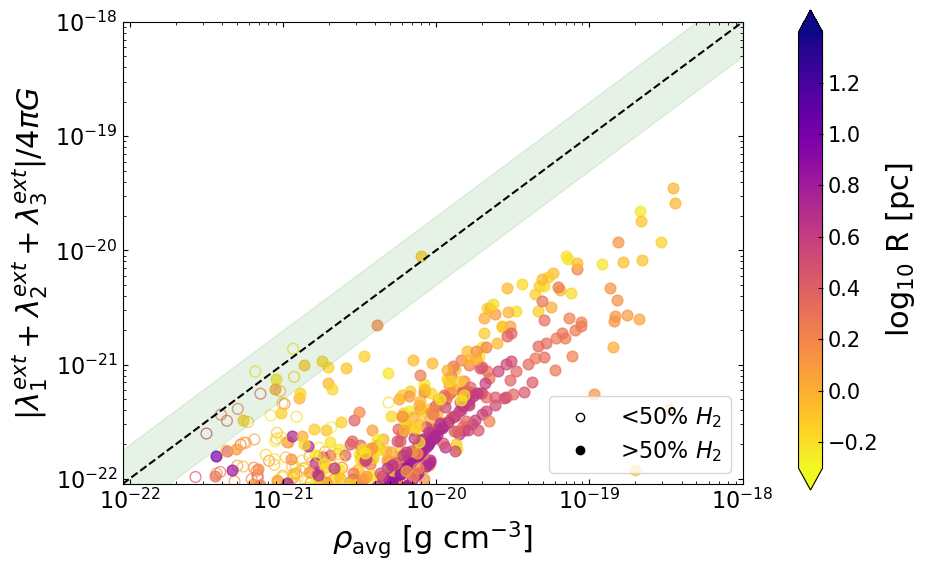}
    \caption{Trace of the averaged total tidal tensor $\langle \mathbf{T}_{\rm tot} \rangle$ (left) and trace of the averaged external tidal tensor $\langle \mathbf{T}_{\rm ext} \rangle$ (right), plotted against the average density, for all cloud sub-structures at $t_{\rm evol}=3.5$~Myr. The dashed line represents a one to one line, and the green shaded region represents a factor of 2 in each direction. The trace of $\langle \mathbf{T}_{\rm tot} \rangle$ well reflects the average density. The trace of $\langle \mathbf{T}_{\rm ext} \rangle$ should ideally be 0, but is typically only ~1\% of the 1:1 line.}
    \label{fig:tidal_tensor_trace}
\end{figure*}
The tidal tensor encodes the gravitational deformation at a given point. An estimated average tidal tensor, averaged over an entire sub-structure, therefore should encode the deformation behaviour of this entire mass element. One of the proofs of concept of this idea is to evaluate how well the trace corroborates to the average density of the structure (see Eq. \ref{eq:trace1}). \par
From Fig. \ref{fig:tidal_tensor_trace}, top panel we see that the expected relation is indeed well retrieved for the total tidal tensor $\langle \mathbf{T}_{\rm tot} \rangle$. The trace for the external tidal tensor $\langle \mathbf{T}_{\rm ext} \rangle$ should ideally be 0, therefore any residual sum reflects the typical error in the estimate of the eigenvalue terms. However, these values are typically only ~1\% of the 1:1 line, and therefore adequately accurate for our calculations.

\bsp	
\label{lastpage}
\end{document}